\documentclass[aps,prb,superscriptaddress,twocolumn,showpacs]{revtex4}
\usepackage{bm}
\usepackage{graphicx}
\usepackage{amsmath}
\usepackage{amsfonts}

\begin{document}

\title{Quantum Monte Carlo study of $S=1/2$ weakly-anisotropic
antiferromagnets on the square lattice}

\author{Alessandro Cuccoli}
%\email{cuccoli@fi.infn.it}
\affiliation{Dipartimento di Fisica dell'Universit\`a di Firenze,
             via G. Sansone 1, I-50019 Sesto Fiorentino (FI), Italy}
\affiliation{Istituto Nazionale per la Fisica della Materia (INFM),
             Unit\`a di Ricerca di Firenze,
             via G. Sansone 1, I-50019 Sesto Fiorentino (FI), Italy}

\author{Tommaso Roscilde}
%\email{roscilde@fi.infn.it}
\affiliation{Dipartimento di Fisica `A. Volta' dell'Universit\`a di Pavia,
             via A. Bassi 6, I-27100 Pavia, Italy}
\affiliation{Istituto Nazionale per la Fisica della Materia (INFM),
             Unit\`a di Ricerca di Pavia,
             via A. Bassi 6, I-27100 Pavia, Italy}

\author{Valerio Tognetti}
%\email{tognetti@fi.infn.it}
\affiliation{Dipartimento di Fisica dell'Universit\`a di Firenze,
             via G. Sansone 1, I-50019 Sesto Fiorentino (FI), Italy}
\affiliation{Istituto Nazionale per la Fisica della Materia (INFM),
             Unit\`a di Ricerca di Firenze,
             via G. Sansone 1, I-50019 Sesto Fiorentino (FI), Italy}

\author{Ruggero Vaia}
%\email{vaia@ifac.cnr.it}
\affiliation{Istituto di Fisica Applicata `Nello Carrara'
             del Consiglio Nazionale delle Ricerche,
             via Panciatichi~56/30, I-50127 Firenze, Italy}
\affiliation{Istituto Nazionale per la Fisica della Materia (INFM),
             Unit\`a di Ricerca di Firenze,
             via G. Sansone 1, I-50019 Sesto Fiorentino (FI), Italy}

\author{Paola Verrucchi}
%\email{verrucchi@fi.infn.it}
\affiliation{Dipartimento di Fisica dell'Universit\`a di Firenze,
             via G. Sansone 1, I-50019 Sesto Fiorentino (FI), Italy}
\affiliation{Istituto Nazionale per la Fisica della Materia (INFM),
             Unit\`a di Ricerca di Firenze,
             via G. Sansone 1, I-50019 Sesto Fiorentino (FI), Italy}

\date{\today}

\begin{abstract}
We study the finite-temperature behaviour of two-dimensional $S=1/2$
Heisenberg antiferromagnets with very weak easy-axis and easy-plane
exchange anisotropies. By means of quantum Monte Carlo simulations,
based on the continuous-time loop and worm algorithm, we obtain a rich
set of data that allows us to draw conclusions about both the existence
and the type of finite-temperature transition expected in the
considered models. We observe that the essential features of the Ising
universality class, as well as those of the
Berezinskii-Kosterlitz-Thouless (BKT) one, are preserved even for
anisotropies as small as 10$^{-3}$ times the exchange integral; such
outcome, being referred to the most quantum case $S=1/2$, rules out the
possibility for quantum fluctuations to destroy the long or quasi-long
range order, whose onset is responsible for the Ising or BKT
transition, no matter how small the anisotropy. Besides this general
issue, we use our results to extract, out of the isotropic component,
the features which are peculiar to weakly anisotropic models, with
particular attention for the temperature region immediately above the
transition. By this analysis we aim to give a handy tool for
understanding the experimental data relative to those real compounds
whose anisotropies are too weak for a qualitative description to
accomplish the goal of singling out the genuinely two-dimensional
critical behaviour.
\end{abstract}

\pacs{75.10.Jm, 75.40.-s, 75.40.Mg, 75.40.Cx}
% 75.10.Jm - Quantized spin models
% 75.40.-s - Critical-point effects, specific heats, short range order
% 75.40.Mg - Numerical simulation studies
% 75.40.Cx - Static properties
% 05.30.-d - Quantum statistical mechanics
% 02.70.Ss - Quantum Monte Carlo methods

\maketitle

\section{Introduction}
In the last few decades the Heisenberg antiferromagnet (HAFM) on the
square lattice has been thoroughly studied by means of several
theoretical, numerical and experimental
techniques~\cite{deJongh90,Manousakis91,Johnston97}. Such research
hands us a picture where three classes of substantially different
models appear, the isotropic, the easy-axis, and the easy-plane
antiferromagnets. The reference models for the two latter classes are
the 2D-Ising and 2D-XY antiferromagnets, defining their respective
universality class and known to display a finite-temperature phase
transition. It is also known that quantum effects, despite being
enhanced by the reduced dimensionality, do not substantially change the
essential features of the above models, not even in the extreme $S=1/2$
case, though evidently renormalizing their thermodynamic properties.
This results from renormalization
group~\cite{DrzewinkskiS89,IrkhinK98-99} and semiclassical
approaches~\cite{CTVV95etc,CRTVV01}, as well as from quantum Monte
Carlo (QMC)
simulations~\cite{QMC-EA,Ding92-EP,Ding92-XY,HaradaK98,Aplesnin98}.

In the above picture, however, there still is a grey area, where our
knowledge is not detailed enough to allow a precise reading of the
experimental data: this is the area of very weak anisotropies and
strong quantum effects, which is of particular interest as most of the
real layered compounds whose magnetic behaviour is properly modeled by
the $S=1/2$ HAFM on the square lattice are characterized by
anisotropies as small as $10^{-3}$ times the exchange integral.  These
compounds exhibit a phase transition at a finite temperature
$T_{\rm{N}}$ which is often too large for the interlayer coupling to be
the unique player, while the idea of a two-dimensional anisotropic
criticality as the trigger of the transition appears well
sound~\cite{HikamiT80,deJongh90a}: such idea is corroborated by the
measured values of some critical exponents~\cite{BramwellH93jap}. The
experimental observation tells us that three-dimensional long-range
order is present below $T_{\rm{N}}$ and that well above $T_{\rm{N}}$ no
trace of anisotropic behaviour is left; it is slightly above
$T_{\rm{N}}$ that one hence expects evidences of genuine 2D anisotropic
behaviour to be detectable. In order to let these experimental
evidences surface out of the sea of the isotropic thermodynamics,
precise numerical data for the $S=1/2$ nearly isotropic HAFM are
needed: it is the purpose of this work to fulfill such need. In
particular, we address the problem of the existence and
characterization of the phase transition, making use of continuous-time
QMC cluster algorithms, which are well suited in the neighbourhood of a
critical point, since they do not suffer from critical slowing-down.
The general resulting picture is that a phase transition is induced by
an arbitrary amount of anisotropy, and that several distinctive
features of the expected universality class can be traced out.

The structure of the paper is as follows: in Section~\ref{s.2DXXZ} the
general properties of the anisotropic models are described, extracting
the most significant difference with respect to the isotropic model. In
Section~\ref{s.defQMCFSE} the QMC methods are presented, the
thermodynamic quantities under investigation are defined, and the
finite-size scaling (FSS) theory used in our analysis is briefly
recalled. The results for nearly isotropic models, as from both FSS
analysis and thermodynamic behaviour, are presented and discussed in
Sections~\ref{s.EAXXZ} and~\ref{s.EPXXZ} for the easy-axis and
easy-plane case, respectively. In Section~\ref{s.phasediag} the
critical-temperature vs anisotropy phase-diagram is discussed.
Eventually, conclusions are drawn in Section~\ref{s.conclusions}.

\section{The two-dimensional XXZ model on the square lattice}
\label{s.2DXXZ}

The XXZ model is defined by the Hamiltonian
 \begin{eqnarray}
 \hat{\cal H} &=& \frac J2\sum_{{\bm{i}},{\bm{d}}}
 \Big[(1{-}\Delta_{\mu})\big(
 \hat{S}_{\bm{i}}^x\hat{S}_{\bm{i}+\bm{d}}^x+
 \hat{S}_{\bm{i}}^y\hat{S}_{\bm{i}+\bm{d}}^y\big)+
\nonumber\\
 && \hspace{31mm}
 +(1{-}\Delta_{\lambda})\,
 \hat{S}_{\bm{i}}^z\hat{S}_{{\bm{i}}+{\bm{d}}}^z\Big]~,
 \label{e.xxzmodel}
 \end{eqnarray}
where ${\bm{i}}=(i_1,i_2)$ runs over the sites of an $L\times L$ square
lattice, ${\bm{d}}$ connects each site to its four nearest neighbours,
$J>0$ is the antiferromagnetic exchange integral, $\Delta_{\mu}$ and
$\Delta_{\lambda}$ are the easy-axis (EA) and easy-plane (EP)
anisotropy parameters, respectively, hereafter getting positive values
smaller than unity. As $J$ sets the energy scale, the dimensionless
temperature $t\equiv{k_{\rm{B}}T/J}$ will be used in the following.

The spin operators  $\hat{S}^\alpha_{\bm{i}}$ ($\alpha=x,y,z$) obey the
{\em su}(2) commutation relations
$[\hat{S}^\alpha_{\bm{i}},\hat{S}^\beta_{\bm{j}}]=
i\varepsilon^{\alpha\beta\gamma}\delta_{\bm{ij}}
\hat{S}^\gamma_{\bm{i}}$ and are such that $|\hat{\bm{S}}|^2=S(S+1)$.

Besides the isotropic model $\Delta_{\lambda}=\Delta_{\mu}=0$,
Eq.~\eqref{e.xxzmodel} defines the EA ($\Delta_{\lambda}=0$,
$0<\Delta_{\mu}\leq 1$) and EP ($\Delta_{\mu}=0$,
$0<\Delta_{\lambda}\leq 1$) magnets, whose respective reference models
are the Ising model ($\Delta_{\mu}=1$) and the XY (also known as XX0)
one ($\Delta_{\lambda}=1$).

What is known about these models can be summarized as follows:

{\it i}) the isotropic model has no finite-temperature transition; its
ground state is ordered for any $S$ and a critical region of divergent
correlations is clearly observed at very low temperatures;

{\it ii}) the EA models exhibit an Ising-like transition at a critical
temperature $t_{_{\rm{I}}}$ which is an increasing function of both
$\Delta_{\mu}$ and $S$; for $S\ge 1$, $t_{_{\rm{I}}}$ is finite for all
anisotropies;

{\it iii}) the EP models exhibit a transition of the
Berezinskii-Kosterlitz-Thouless (BKT) type, at a critical temperature
$t_{_{\rm{BKT}}}$ which is an increasing function of both
$\Delta_{\mu}$ and $S$; for $S\ge 1$, $t_{_{\rm{BKT}}}$ is finite for
all anisotropies.

Some of the above statements are rigorously proved, others come from
the combination of theoretical, numerical and experimental results.

More precisely, for what concerns ground state properties, an
antiferromagnetically ordered ground state has been exactly
proved~\cite{WischmannMH91,KuboK88} for $S=1/2$ only in the case of
strong anisotropy, $\Delta_{\lambda}>0.88$ and $\Delta_{\mu}>0.32$. For
weaker anisotropies clear numerical evidence exists of an ordered
ground state down to the isotropic limit, as coming from QMC and exact
diagonalization studies~\cite{groundstate}. As for the
finite-temperature properties, according to renormalization group
analysis~\cite{Khokhlacev76,IrkhinK98-99}, the critical temperature
vanishes in the isotropic limit as
$t_{\rm{I,BKT}}\sim[\ln({\rm{const.}}/\Delta_{\rm{\mu,\lambda}})]^{-1}$.
In the EA case the existence of the transition for arbitrary
$\Delta_{\mu}$ has been rigorously proved only in the classical limit
$S\to\infty$, while in the quantum case the proof is restricted to a
spin-dependent finite amount of anisotropy~\cite{FroelichL77} . In the
EP case the situation is even less clear, since also in the classical
limit a rigorous proof~\cite{Dunlop85} of the existence of a transition
is limited to the case $\Delta_{\lambda}\gtrsim 0.553$, while no
rigorous proof exists in the quantum case.

In the $S=1/2$ case, evidences of a phase transition for anisotropies
as small as $\Delta_{\mu}=0.01$ in the EA case~\cite{QMC-EA,Aplesnin98}
and $\Delta_{\lambda}=0.02$ in the EP case~\cite{Ding92-EP} were
suggested by previous QMC approaches; however, these computations
employed local algorithms which cannot easily access the critical
region and a rigorous FSS analysis could not be performed. The
situation is still unclear also because recent works based on
real-space renormalization-group~\cite{BrancoRdS00,RicardodeSousaA00}
predict the existence of a critical value of the anisotropy in the EA
case ($\Delta_{\mu}^{({\rm{c}})}\approx0.2$) below which the transition
would be destroyed by quantum fluctuations. In this work we hence
consider $S=1/2$ and four nearly isotropic systems, two EA,
$\Delta_{\mu}=0.01$ and $\Delta_{\mu}=0.001$, and two EP ones,
$\Delta_{\lambda}=0.02$ and $\Delta_{\lambda}=0.001$.

\section{Quantum Monte Carlo, observables and finite-size effects}
\label{s.defQMCFSE}

\subsection{Quantum Monte Carlo method: continuous-time algorithms}

As usually done in the existing literature on QMC, in this section and
in appendices~\ref{a.worm} and~\ref{a.helicity} we employ the notation
\begin{equation}
 J^{XY}\equiv J(1-\Delta_{\mu})~,
 ~~~~~ J^{Z}\equiv J(1-\Delta_{\lambda})~,
\end{equation}
and $\beta\equiv1/k_{\rm{B}}T$.

The QMC method for the $S=1/2$ XXZ model is based on the Trotter-Suzuki
decomposition of the partition function, which can be approximated by
the following expression~\cite{Suzuki76-BarmaS78}:
\begin{equation}
 {\cal Z(\beta)}\equiv \text{Tr}~e^{-\beta {\hat {\cal H}}}
 \approx \sum_{\cal S} \prod_n w_{{\rm{p}}_{n,{\cal S}}}(\Delta \tau)~,
\label{Trotter}
\end{equation}
where $w_{\rm{p}}$ represents the amplitude of propagation of a pair of
nearest-neighbour spins from a configuration
$|\sigma_{\bm{i}},\sigma_{\bm{j}}\rangle$ to
$|\sigma_{\bm{i}}',\sigma_{\bm{j}}'\rangle$ in the (imaginary-) time
step $\Delta \tau = \beta/M$, $M$ being the Trotter number 
and $|\{\sigma_{\bm{i}}\}\rangle$
($\sigma_{\bm{i}}=\pm 1/2$) the basis set diagonalizing the
$\hat{S}^z_{\bm{i}}$ operator.  The two bond configurations define a
space-time {\it plaquette} configuration
${\rm{p}}=\{\sigma_{\bm{i}},\sigma_{\bm{j}};
\sigma_{\bm{i}}',\sigma_{\bm{j}}'\}$,
so that $w_{\rm{p}}$ can be seen also as the weight of a given
plaquette configuration ${\rm{p}}$. The index $n$ runs over all
plaquettes on the space-time lattice, and the index ${\cal S}$ runs
over all configurations of the system. At each time step plaquettes are
defined on different groups of bonds
$\langle{{\bm{i}}{\bm{j}}}\rangle$, so that all bonds involved in the
propagation at the same time step do not share any spin; moreover, each
plaquette shares its corner spins with two plaquettes on the previous
and two on the subsequent time step. The expression~\eqref{Trotter}
becomes exact in the limit $M\to\infty$.

\begin{figure}
\includegraphics[bbllx=5mm,bblly=4mm,bburx=164mm,bbury=128mm,%
     width=70mm,angle=0]{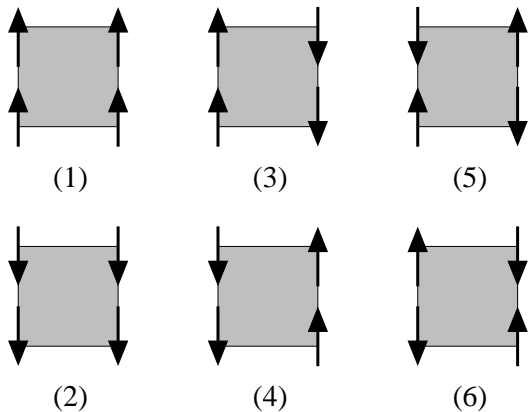}
 \caption{\label{plaquettes}
Plaquette configurations with non-vanishing weights in the $S=1/2$ XXZ
model. The vertical axis is the imaginary-time direction.}
\end{figure}

In the case of the XXZ model, only the 6 plaquette configurations shown
in Fig.~\ref{plaquettes} have non-zero weights $w_{\rm{p}}$, whose
expansion to first non trivial order in $\Delta\tau$
are~\cite{Evertz,nota}:
\begin{eqnarray}
 w_1 = w_2 \approx 1 - \frac{J^Z}{4}\Delta\tau
\nonumber \\
 w_3 = w_4 \approx 1 + \frac{J^Z}{4}\Delta\tau
\nonumber \\
 w_5 = w_6 \approx \frac{J^{XY}}{2}\Delta\tau~.
\end{eqnarray}
Plaquettes 1,~2,~3, and 4 propagate the state of the spin pair
unchanged, while plaquettes~5 and~6 introduce an exchange of state for
the interacting spins, hereafter denoted as {\it kink}.

Any standard Monte Carlo (MC) method can be applied for numerically
evaluating the partition function~\eqref{Trotter}; in principle one
could do single local Metropolis moves which update clusters of
plaquettes sharing corner spins in order to move to configurations with
non-zero weight~\cite{CullenL83}.  This approach is however
inconvenient, being affected by the well-known critical slowing-down.
Moreover, in order to control the Trotter-approximation error, it is
necessary to repeat the simulation for increasing Trotter number. In
recent years efficient cluster algorithms were proposed to overcome the
problem of critical slowing down in QMC simulations. In the context of
quantum spin systems, the {\it loop-cluster algorithm} and the {\it
worm algorithm} have revealed very effective; moreover, they can be
also formulated in continuous imaginary time, thus removing completely
the Trotter-approximation error.

\subsubsection{Loop algorithm}

\begin{figure}
\includegraphics[bbllx=7mm,bblly=31mm,bburx=201mm,bbury=76mm,%
     width=80mm,angle=0]{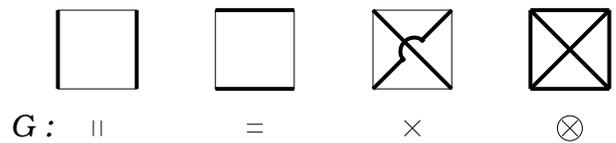}
 \caption{\label{breakups}
Breakup decompositions of a single plaquette in the $S=1/2$ XXZ model;
thick lines join grouped spins.}
\end{figure}

One can completely eliminate the critical slowing-down by introducing
the so-called {\it loop-cluster algorithm}~\cite{Evertz}, which is the
quantum analog of the Swendsen-Wang~\cite{SwendsenW87} and
Wolff~\cite{Wolff89} cluster algorithms introduced for classical spin
systems. Within the multi-cluster approach~\cite{HaradaK98} (analogue
to the Swendsen-Wang scheme) the loop algorithm consists of
probabilistically assigning to each plaquette a {\it break-up
decomposition} (or graph) $G$, i.e., a way of grouping its spins in
subgroups, so that the grouped spins can be flipped all at once
bringing the plaquette into a configuration with non-vanishing weight;
in the case of the XXZ model the above condition allows only grouping
of spins into pairs ({\it non-freezing} breakups) or all together ({\it
freezing} breakup), as shown in Fig.~\ref{breakups}.
Assigning a breakup $G$ to each plaquette, one univocally defines a
breakup decomposition ${\cal G} = \{G_n\}$ of the whole configuration
${\cal S} = \{\sigma_{i,\tau}\}$ into loops.  Different breakup
decompositions $G$ are assigned to each plaquette configuration
${\rm{p}}$ according to weights $w({\rm{p}},G)$ obeying the general sum
rule $w_{\rm{p}}=\sum_{G}w({\rm{p}},G)$.  Since the XXZ model is
symmetric under time reversal, it is straightforward to require that
plaquettes connected by time-reversal must have the same weights, i.e.,
$w(1,G) = w(2,G)$, $w(3,G) = w(4,G)$ and $w(5,G) = w(6,G)$.  These
weights must be positive and obey the detailed balance condition.  The
presence of frozen plaquettes leads to the generation of very big
loops, whose flip is expected to be less and less effective in
generating successive uncorrelated configurations; therefore, a general
strategy to be followed is that of minimizing the probability of frozen
plaquettes to appear. If one sets $w({\rm{p}}, \otimes)=0$ for each p,
a solution with positive breakup weights is obtained only in the case
$J^{XY} \geq J^{Z}$, i.e., in the EP and isotropic case. When $J^{Z} >
J^{XY}$, i.e., in the EA case, one must allow for freezing of at least
a single plaquette configuration to achieve positiveness. In
particular, in the antiferromagnetic case, it is
$w(3,\otimes)\equiv{w(4,\otimes)}$ that must be different from zero in
order to have positive weights. By further imposing independence of the
breakup decomposition from the plaquette configuration, i.e.,
\begin{eqnarray}
 w(1,\parallel) = w(3,\parallel)
 \hspace{5mm}  w(1,\times) = w(5,\times)
\nonumber \\
 w(3,=) = w(5,=) \hspace{1.5cm}~,
\end{eqnarray}
one gets that, within the same loop decomposition ${\cal G}$, all of
the configurations ${\cal S'}$ obtained from the starting one ${\cal
S}$ by flipping some of the loops, are reached with the same
probability, i.e., each loop is flipped independently of all the other
ones.

The complete set of non-zero breakup weights used in our calculation
thus reads~\cite{Evertz} (to first order in $\Delta\tau$):
\begin{itemize}
\item {Easy-plane and isotropic case} ($J_x \geq J_z$):
\begin{eqnarray}
 w(1,\parallel) & = & 1 - \frac{J^{XY}}{4} \Delta\tau
\nonumber \\
 w(1,\times)    & = & \frac{J^{XY}-J^Z}{4} \Delta\tau
\nonumber \\
 w(3,=)  & = & \frac{J^{XY}+J^Z}{4} \Delta\tau~;
\label{EP-breakup}\end{eqnarray}
\item {Easy-axis case} ($J_x < J_z$):
\begin{eqnarray}
 w(1,\parallel) & = & 1 - \frac{J^{Z}}{4} \Delta\tau
\nonumber \\
 w(3,=) & = & \frac{J^{XY}}{2} \Delta\tau
\nonumber \\
 w(3,\otimes) & = & \frac{J^Z-J^{XY}}{2} \Delta\tau~.
\label{EA-breakup}
\end{eqnarray}
\end{itemize}

The fact that freezing is required in the EA case can be physically
interpreted as a reflection of the enforced antiferromagnetic
correlations of the $z$-component of the spins: as a matter of fact,
freezing of plaquettes of type 3 and 4 implies preservation of
nearest-neighbour antiferromagnetic correlation along the MC flow,
i.e., antiferromagnetic configurations are more ``resistant'' to MC
fluctuations, with respect to ferromagnetic ones.

In the limit of continuous imaginary time~\cite{BeardW96},
$\Delta\tau\to0$, plaquettes with no {\it kink}, i.e., of types 1 (2)
and 3 (4), acquire unitary weight, while plaquettes with a kink, i.e.,
of type 5 (6), get an infinitesimal weight, still keeping a finite
weight per unit time $\omega_{\rm{p}}=\lim_{\Delta\tau\to
0}w_{\rm{p}}/\Delta\tau$; therefore kink-bearing plaquettes must be
regarded as Poissonian events in the imaginary-time evolution of each
pair of interacting spins. At the same time the breakup decomposition
creating no kink in imaginary time evolution, acquires a unitary
weight, while all the other breakups acquire infinitesimal weight,
still keeping a finite weight per unit time $\omega({\rm{p}},G)$. In
the case of plaquette 5 (6), since the breakup weights have to be
normalized to the plaquette weights, they become finite probabilities
\begin{equation}
p(5,=) =  \frac{1}{2} \left ( 1 + \frac{J^Z}{J^{XY}} \right )
\hspace{.5cm}
p(5,\times) = 1 - p(5,=)
\end{equation}
in the EP case, and $p(5,=) = 1$, $p(5,\times) = 0$ in the EA case.

The algorithm then proceeds as follows:
\begin{itemize}
\item distribute breakups $\times$, $=$ and $\otimes$ on
the continuous segments (= infinite sequences of infinitesimal
plaquettes 1(2) and 3(4)) along the imaginary-time evolution of each
pair of interacting spins, according to the Poisson distribution having
as parameter $\beta\,\omega({\rm{p}},G)$; for each kink in the
propagation, choose a breakup with probabilities $p(5,G)$.
\item reconstruct the loops defined by the decomposition
of each infinitesimal plaquette;
\item decide whether to flip each loop independently
with probability one half.
\end{itemize}
The above procedure (multi-cluster update) represents a single MC step
in our code. We have generally performed $10^4$ MC steps for
thermalization for each value of the temperature, and $1\div 1.5\cdot
10^5$ MC steps for evaluation of thermodynamic observables. The
algorithm is very efficient in both the EA and the EP case, with
autocorrelation times which always remain around unity for all the
lattice sizes $L$ we considered, i.e., $L=$16, 32, 64, 128, 200.  The
autocorrelation time $\tau_{\rm{c}}$ has been estimated by the 
{\it blocking technique}\cite{Janke02} as:
\begin{equation}
\tau_{\rm{c}} = \frac {N_{\rm{b}}}{2} ~\frac{\sigma_X^2}{\sigma_x^2}
\end{equation}
where $\sigma_x^2$ is the variance of the time-series $\{x_i\}$
($i=1,...,N_{\rm{steps}}\,{=}\,n_{\rm{b}}N_{\rm{b}}$) produced for the
variable $x$, while $\sigma_X^2$ is the variance of the block variable
$X_j=N_{\rm{b}}^{-1}\sum_{i=(j-1)*N_{\rm{b}}+1}^{j*N_{\rm{b}}}~x_i$
($j=1,...,n_{\rm{b}}$), with $N_{\rm{b}} \gg \tau$ for the estimate to
be sensible.

The introduction of freezing breakups has the effect of making a loop
branch into subloops; as the flip of large branched clusters is less
effective in generating successive uncorrelated configurations, the
inclusion of freezing is generally thought to lower the efficiency of
the loop algorithm~\cite{Evertz}, although no direct evidence of such
conclusion exists. Nevertheless, for the EA anisotropies we consider,
no significant loss of efficiency (i.e., increase in correlation time)
is observed. Moreover, in the Ising limit, the above algorithm
reproduces the Swendsen-Wang algorithm for the Ising model, which is
known not to suffer from slowing down; therefore efficiency is possibly
lost only for intermediate values of the anisotropy, which is not the
case we are interested in.

We have implemented improved estimators~\cite{Wolff89,Broweretal98} for
all the quantities of interest. A separate, more careful analysis is
needed in the case of off-diagonal observables, whose most general
bilinear example may be
$\big\langle\hat{S}^+_{\bm{i}}(\tau)\hat{S}^-_{\bm{j}}(\tau')\big\rangle$.
In absence of freezing, the improved estimator simply
reads~\cite{Broweretal98,AlvarezG00}: $\big( \hat{S}^+_{\bm{i}}(\tau)
\hat{S}^-_{\bm{j}} (\tau')\big)_{\rm{impr}}= 1$, if $(\bm{i},\tau)$ and
$(\bm{j},\tau')$ belong to the same loop, $=0$ otherwise. This result
can be understood by observing that the quantity we are considering may
be seen as the ratio between the partition function of the system with
density operator $\hat{S}^+_{\bm{i}}(\tau) \hat{S}^-_{\bm{j}} (\tau')
\exp(-\beta \hat{\cal H})$ and the partition function of the original
model considered, with density operator $\exp(-\beta
\hat{\cal H})$. The final outcome for the estimator given above thus
follows from the observation that in absence of freezing, i.e., in the
isotropic or EP class, according to Eqs.~\eqref{EP-breakup} the weight
assigned to a loop decomposition is actually independent of the
configuration onto which it is defined, since each kind of breakup has
the same weight on every plaquette configuration where its weight is
non-zero; therefore the same loop decomposition is reached with equal
probability in the original and in the modified system, with the
additional constraint, in the case of the modified system, that the two
points $(\bm{i},\tau)$ and $(\bm{j},\tau')$ must belong to the same
loop.

When freezing is present, the above argument breaks down, since only
one plaquette configuration admits freezing and the constraint of
having $(\bm{i},\tau)$ and $(\bm{j},\tau')$ on the same loop is no
longer sufficient to have a non-zero contribution to the estimator. In
principle it is possible to define the estimator for off-diagonal
observables even in case of freezing; however, not only its
implementation is highly non trivial from the point of view of
programming, but its evaluation would also consume a considerable
amount of computational time. As a matter of fact, for each pair of
points $(\bm{i},\tau)$, $(\bm{j},\tau')$ belonging to the same branch
of a loop, one should look for freezing breakups by checking all the
loop segments between the two points, an operation that, being the
loops structure involved, becomes very costly at low temperatures. We
have therefore renounced to implement such estimators in case of
freezing. To have a complete picture of the thermodynamics of the
system in the EA case we have then resorted to a different (and
generally less efficient) QMC scheme, within which the calculation of
the off-diagonal observables in the EA case is relatively
straightforward, i.e., to the so-called worm algorithm.

\subsubsection{Worm algorithm}

The {\it worm} algorithm represents an alternative way to overcome the
problem of critical slowing down in QMC simulations. The original idea
of the algorithm can be found in Ref.~\onlinecite{worms}, but we here
formulate the algorithm in a different way, so that it appears as a
direct generalization of the loop algorithm; our formulation is more
directly related with the so-called "operator-loop update" introduced
in the framework of the stochastic series expansion~\cite{Sandvik99}.

The worm algorithm starts by choosing a point at random in space-time,
inserting two discontinuities in the local imaginary-time evolution,
and then keeping one fixed (the "tail" of the worm) while letting the
other (the "head" of the worm) freely travel through the lattice. The
single-worm update ends when the head happens to "eat" the tail (the
worm closes), so that the isolated discontinuities disappear and the
system is led to a new configuration having non-zero weight.  All the
segments of imaginary-time evolution touched by the worm's head have to
be flipped, i.e., the worm's head performs a real-time update of the
system. Its motion conventionally goes forward (backward) in imaginary
time while updating segments with up (down) spins, and it is ruled by
detailed balance condition, to be locally satisfied on each
(infinitesimal) plaquette it touches. In the context of the worm
algorithm one has to abandon the concept of plaquette breakup, and the
decision about how a plaquette is passed through by the worm must be
taken from time to time. One of the distinctive feature of the
algorithm is that the worm can pass through the same plaquette many
times, each time in a different way, depending on the local
configuration left at its previous passage.

General detailed balance conditions for the single plaquette update
when the worm's head passes through it, flipping two spins, read
\begin{eqnarray}
 w_1 ~ p(1{\to}3) &=& w_3 ~ p(3{\to}1)
\nonumber \\
 w_1 ~ p(1{\to}5) &=& w_5 ~ p(5{\to}1)
\nonumber \\
 w_3 ~ p(3{\to}5) &=& w_5 ~ p(5{\to}3)~,
\end{eqnarray}
where we have already introduced the time- and space-reversal
symmetries, so that here "1" means 1 or 2, "3" means 3 or 4, and "5"
means 5 or 6, depending on the way the worm's head travels through the
plaquette.  Moreover, the transition probabilities must satisfy the sum
rules
\begin{eqnarray}
 p(1{\to}3) + p(1{\to}5) &=& 1
\nonumber \\
 p(3{\to}1) + p(3{\to}5) &=& 1
\nonumber \\
 p(5{\to}1) + p(5{\to}3) &=& 1~.
\label{sumrules}
\end{eqnarray}
Moving to the continuous-time limit, we express the transition
probabilities from a plaquette with a non-vanishing weight to another
plaquette, as
\begin{eqnarray}
 p(1{\to}3) &=& 1 + \pi(1{\to}3)\,\Delta\tau
\nonumber \\
 p(3{\to}1) &=& 1 + \pi(3{\to}1)\,\Delta\tau
\nonumber \\
 p(1{\to}5) &=& \pi(1{\to}5)\,\Delta\tau
\nonumber \\
 p(3{\to}5) &=& \pi(3{\to}5)\,\Delta\tau ~,
\end{eqnarray}
where $\pi(1{\to}5)$ and $\pi(3{\to}5)$ have the meaning of (positive)
transition probabilities per unit imaginary time, while $\pi(1{\to}3)$
and $\pi(3{\to}1)$ are (negative) corrections to the transition
probabilities among plaquettes taking non-vanishing weights; these
corrections arise from the poissonian occurrence of kinks in
imaginary-time evolution. With the above parameterization of the
transition probabilities, the first two of the sum
rules~\eqref{sumrules} take the form
\begin{eqnarray}
 \pi(1{\to}3) &+& \pi(1{\to}5) = 0
\nonumber \\
 \pi(3{\to}1) &+& \pi(3{\to}5) = 0 ~,
\end{eqnarray}
while the third remains unchanged, as $p(5{\to}1)$ and $p(5{\to}3)$
keep their meaning of dimensionless probabilities for the different
ways the worm's head can pass through a kink in imaginary-time
evolution. The set of detailed balance equations in the continuous-time
limit takes the form
\begin{eqnarray}
 \pi(1{\to}3) - \pi(3{\to}1) &=& \frac{J^Z}{2}
\nonumber \\
 \pi(1{\to}5) &=& \frac{J^{XY}}{2} ~ p(5{\to}1)
\nonumber \\
 \pi(3{\to}5) &=& \frac{J^{XY}}{2} ~ p(5{\to}3) ~.
\end{eqnarray}
Together with the sum rules, they give as unique solution the following
set of transition probabilities:
\begin{eqnarray}
 \pi(1{\to}5) &=& \frac{J^{XY}-J^{Z}}{4}
\nonumber \\
 \pi(3{\to}5) &=& \frac{J^{XY}+J^{Z}}{4}
\nonumber \\
 p(5{\to}3) &=& \frac12\bigg(1-\frac{J^{Z}}{J^{XY}}\bigg)~.
\label{e.paip}
\end{eqnarray}
It is immediate to see that this solution is equivalent to the set of
breakup weights~\eqref{EP-breakup}, which means that,  at this level,
the worm algorithm is nothing but the Wolff-type (single-cluster)
version of the loop algorithm. However, as seen in the loop algorithm,
in the EA case transition probabilities are not always positive, and
some other transition mechanism must be invoked to overcome this
problem. As seen before, the remedy in the case of the loop algorithm
was to allow for branching of the loops; if one hence allows for
branching also in the worm algorithm, the single-cluster version of the
loop algorithm for the EA case is obtained.

A different strategy can be adopted in the case of the worm algorithm
by introducing a new type of motion, named {\it bouncing}, where the
worm's head, when attempting to update a plaquette, is bounced off and
hence forced to locally trace back its route. From the physical point
of view, the existence of a bounce mechanism protects some plaquettes
from being updated, possibly those plaquettes containing local spin
configurations which give a relevant contribution to the thermodynamics
of the system; such local configurations are preserved along the MC
flow and the effect of bouncing is seen to be very similar to that of
freezing in the context of the loop algorithm.  In the case of the EA
antiferromagnet, the most relevant local configurations are those
containing antiferromagnetic correlations of the $z$-components, i.e.,
in terms of plaquettes, p=3,4.  Therefore, we allow for bounce on these
plaquette configurations, introducing a finite bounce probability
$p(3,b) =
\pi(3,b) \Delta\tau$ which has to be accounted for in the sum rule
\begin{equation}
\pi(3{\to}1) + \pi(3{\to}5) + \pi(3,b) = 0 ~.
\label{e.pai}
\end{equation}
The detailed balance condition for the bounce probability is trivial,
reading $p(3,b)~w(3) = p(3,b)~w(3)$. Eqs.~\eqref{e.paip}
and~\eqref{e.pai} form an underdimensioned set and $\pi(3,b)$ can be
hence chosen arbitrarily, with the only constraint of positive
transition probabilities. As in the case of freezing, it is highly
convenient to minimize the bounce probability: when the worm's head
bounces, part of its update operations are lost as it locally traces
back its way, so that the efficiency in updating the configuration,
keeping the number of elementary update operations fixed, is lowered.
The following solution for the transition probabilities, minimizing the
bounce probability, is found
\begin{eqnarray}
 \pi(1{\to}5) &=& 0
\nonumber \\
 \pi(3{\to}5) &=& \frac{1}{2}~J^{XY}
\nonumber \\
 \pi(3,b) &=& \frac{1}{2}\big(J^Z-J^{XY}\big)
\nonumber \\
 p(5{\to}1) &=& 1~.
\label{bounce-set}
\end{eqnarray}
The worm algorithm with the bounce process is a {\it pure-quantum}
cluster algorithm: in the Ising model, which is a substantially
classical statistical model, the algorithm loses its cluster nature,
since only bounce processes survive, thus confining the worm on a
single site. In this case the worm algorithm reproduces the single-flip
Metropolis algorithm, as shown in Appendix~\ref{a.worm}.

As in the case of the loop algorithm, each of our simulations consists
of $10^4$ MC steps for thermalization and of $1\div 1.5*10^5$ MC steps
for evaluation of thermodynamic observables. During thermalization, the
number of worms to be produced at each step is adjusted so that the
total length of the worms in the imaginary-time direction roughly
equals the size of the (D+1)-dimensional lattice, $L^2*\beta$; this
number is then kept fixed during the measurement phase. In this way,
autocorrelation times of the order of unity are achieved for all values
of the EA anisotropy considered. At variance with the loop algorithm,
the efficiency is here expected to drastically decrease as the
anisotropy increases, given that, as the model moves toward the Ising
limit,  the cluster algorithm transforms into the local Metropolis
algorithm; however, the case of strong anisotropy is not of our
interest here.

As mentioned in the previous section, the estimator for bilinear
off-diagonal quantities like
$\hat{S}^+_{\bm{i}}(\tau)\hat{S}^-_{\bm{j}}(\tau')$ can be thought of
as a partition function for a modified model, in which two spin
discontinuities are inserted in the system configuration at the points
$(\bm{i},\tau)$, $(\bm{j},\tau')$.  Now it becomes clear that
configurations giving a non-zero contribution to such a partition
function are generated during the worm update whenever the
discontinuities associated to the head and tail of the worm coincide
with the above points, both in the EA and in the EP case. Therefore the
off-diagonal observables are measured on-the-fly during the motion of
worm's head~\cite{worms}, and each worm update produces a statistics
for the estimators which grows linearly with the length of the worm. As
in the case of the loop algorithm, this kind of estimators give to the
worm geometry a physical meaning: the further the worm head travels
away from its tail, the larger will be the off-diagonal correlations.
On the other hand, improved estimators are not defined for diagonal
quantities in the EA case; to this respect, worm and loop algorithms
are seen to be exactly complementary.

As final remarks, we mention that the worm algorithm retains its full
efficiency also in presence of a uniform magnetic field applied to the
spins, while the loop algorithm is known to exponentially slow down as
the field is increased and/or the temperature is
lowered~\cite{Onishietal99}. Details of the worm algorithm in a finite
field are not relevant to the present work and will be given 
elsewhere. Independently of us, Sylju{\aa}sen and 
Sandvik~\cite{SyljuasenS2002}
have recently developed a very similar ({\em{directed loop}}) 
algorithm, in the framework of both stochastic 
series expansion and path-integral Monte Carlo.

\subsection{Thermodynamic quantities}
\label{ss.defs}

We briefly report here the definition of the relevant thermodynamic
quantities measured in our QMC study, together with their respective
estimators. The MC average of the estimator will be hereafter denoted
as $\langle...\rangle_{_{\rm{MC}}}$.

The internal energy $\langle \hat{\cal H} \rangle$ is estimated as the
MC average of
\begin{equation}
\frac{1}{2\beta}
\sum_{{\bm{i}},{\bm{d}}} \int_0^{\beta} d\tau ~
\phi_{{\bm{i}},{\bm{d}}}(\tau)\equiv E~,
\end{equation}
where  $\phi_{{\bm{i}},{\bm{d}}}(\tau)$ takes the value $- J^Z/4$  if
at imaginary time $\tau$ there is an infinitesimal plaquette
configuration of type 1 (2), $J^Z/4$ if of type 3 (4),  $-\delta(\tau)$
if of type 5 or (6). This corresponds to the continuous-time limit of
the energy estimator as defined in Ref.~\onlinecite{CullenL83}.

The specific heat $c\equiv\beta^2(\langle\hat{\cal{H}}^2\rangle-
\langle\hat{\cal{H}}\rangle^2)/L^2$ is estimated from energy fluctuations as
\begin{equation}
 \frac1{L^2}~
 \Big(\big\langle \beta^2 E^2 - N_{\rm{kinks}}\big\rangle_{_{\rm{MC}}}
 -\beta^2 \langle E \rangle_{_{\rm{MC}}}^2 \Big)~,
\end{equation}
where $N_{\rm{kinks}}$ is the number of kinks present in each generated
configuration. The variance of the specific heat has been estimated via
binning analysis of the time series related to the energy estimator and
the kink number.

The staggered magnetization $M_{_{\rm{s}}}\equiv(-1)^{\bm{i}}
 \langle\hat{S}^z_{\bm{i}}\rangle$ is estimated as the MC average of
\begin{equation}
\frac{1}{L^2} \sum_{\bm{i}} (-1)^{\bm{i}}
 \sigma^z_{\bm{i}}\equiv m_{_{\rm{s}}}~.
\label{e.qmc.ms}
\end{equation}

The spin-spin correlation function is
\begin{equation}
 C^{\alpha\alpha}(\bm r)=\frac{1}{\beta^2}
 \int_0^{\beta} d\tau d\tau' \big\langle\hat{S}^{\alpha}_{\bm{i}}(\tau)
 \hat{S}^{\alpha}_{{\bm{i}}+{\bm r}}(\tau') \big\rangle~ f(\tau,\tau')\,,
\label{e.defs.cr}
\end{equation}
where $f(\tau,\tau')=\beta~\delta(\tau-\tau')$ in the equal-time
correlator (ET) and $f(\tau,\tau')=1$ in the time-averaged (TA) one. In
both cases the numerical calculation of the correlation function takes
advantage of the existence of the improved estimator defined in the
previous subsection.

The generalized susceptibility is
\begin{equation}
 \chi^{\alpha \alpha}({\bm q}) = \beta
 \sum_{\bm r}~e^{i {\bm q}\cdot {\bm r}}
~ C^{\alpha\alpha}({\bm r})~;
\label{e.defs.chiq}
\end{equation}
the time-averaged susceptibility corresponds to the thermodynamic
definition (second derivative of the free energy) while the equal-time
one corresponds to $\beta*S(\bm q)$, where $S(\bm q)$ is the static
structure factor as measured, e.g., in neutron scattering experiments.
From the general definition above follow those of the uniform
susceptibility
\begin{equation}
 \chi^{\alpha\alpha}_{_{\rm{u}}}=\chi^{\alpha\alpha}({\bm q}{=}0)~,
\label{e.defs.chiu}
\end{equation}
and of the staggered one
\begin{equation}
 \chi^{\alpha\alpha}_{_{\rm{s}}}
 =\chi^{\alpha\alpha}\big({\bm q}{=}(\pi,\pi)\big) ~.
\label{e.defs.chis}
\end{equation}
Susceptibilities and correlation functions have been measured both
along the $z$-axis ($C^{zz}$, $\chi^{zz}$) and in the $xy$-plane
($C^{xx}=C^{yy}$, $\chi^{xx}=\chi^{yy}$); in the EA case, the latter
have been evaluated by means of the worm algorithm.  In what follows,
we will show and comment our data relative to the uniform TA
susceptibility and to the staggered ET susceptibility, being such
quantities the more relevant ones from the experimental point of view.

The correlation length $\xi^{\alpha\alpha}$ is defined {\it via} the
long-distance exponential decay of the staggered correlation function,
$(-1)^{\bm r} C^{\alpha\alpha}({\bm r}) \sim
\exp(-r/\xi^{\alpha\alpha})~~ (r\to\infty)$.  A direct
estimate $\xi^{\alpha\alpha}_{\rm{fit}}$ of the correlation length may
be hence found by fitting the long-distance behaviour of
$C^{\alpha\alpha}({\bm r})$ with a model-dependent function, as
discussed in the following sections.  Such procedure, however, is
strongly dependent on the quality and stability of the fit, and does
not always lead to a univocally defined result in case of a finite-size
system in presence of a phase transition, i.e., of a diverging
correlation length.  An alternative strategy, which we have also used,
is offered by the so-called second moment
definition~\cite{Cooperetal82}
\begin{equation}
\xi_{2}^{\alpha\alpha} =
\frac{L}{2 \pi}
\sqrt{\frac{\chi^{\alpha \alpha} (\pi,\pi)}
{\chi^{\alpha \alpha} (\pi+2\pi/L,\pi)} - 1}~,
\label{secondmom}
\end{equation}
which can be directly extracted by the simulation data, supplemented by
a binning analysis of susceptibility time series in order to estimate
the variance.

Another relevant observable, in the EP case, is the helicity modulus
$\Upsilon$, which is a measure of the response of the system to the
application of a twist $\Phi$ in the boundary condition along a given
direction:
\begin{equation}
\Upsilon \equiv \frac{1}{J^{XY}L^2} \left[\frac{\partial^2 F(\phi)}
{\partial\phi^2}
\right]_{\phi=0}~,
\label{thermohelix}
\end{equation}
where $\phi=\Phi/L$. In Appendix~\ref{a.helicity} we show that,
starting from the above definition as explicitly written in terms of
spin operators, the estimator of the helicity modulus of the $S=1/2$
XXZ EP model reads
\begin{equation}
\Upsilon = \frac{t}{2} |{\bf W}|^2~,
\label{winding}
\end{equation}
where ${\bf W}=(W_1,W_2)$, $W_{\rm{1(2)}}$ being the total winding
number of spin paths (paths traced by a fixed spin configuration, up or
down) in the 1(2) lattice direction. Remarkably, this estimator is
directly related with that of the superfluid density of bosonic
systems~\cite{PollockC87}. An efficient improved version of the
estimator~\eqref{winding} has been introduced by Harada and
Kawashima~\cite{HaradaK98} in the context of the loop algorithm, and is
the one employed in this work.

\subsection{Finite-size scaling}
\label{ss.FSS}

A FSS analysis~\cite{Barber83} can give strong indications on the
existence of a phase transition at some temperature $t_{\rm{c}}$,
possibly leading to a full characterization of its universality class.
The simplest evidence that a transition occurs is found when, for
increasing lattice size, the order parameter scales to a finite value
below a certain temperature, indicating that a non-zero order parameter
develops in the thermodynamic limit.

In the case of second-order phase transitions, the
Ansatz~\cite{Barber83} for the scaling behaviour of a generic
finite-size thermodynamic quantity $A_L(t)$ in the neighbourhood of the
critical point reads
\begin{equation}
 A_L(t)~\sim~ L^{\rho/\nu}~F_A\big[L^{1/\nu}\,(t-t_{\rm{c}})\big]~,
\label{e.fss.At}
\end{equation}
where $\rho$ is the critical exponent of $A\equiv{A_\infty}$, i.e.,
$A(t{\to}t_{\rm{c}})\sim|t-t_{\rm{c}}|^{-\rho}$, $\nu$ is the exponent
for the correlation length, while $F_A$ is the universal scaling
function.  At the critical point Eq.~\eqref{e.fss.At} implies
$A_L(t_{\rm{c}})~\sim~ L^{\rho/\nu}$.  In the case of $\xi$ this means
linear scaling at criticality, without any assumption on the
universality class; therefore, looking for the temperature at which a
properly defined\cite{Caraccioloetal01} $\xi_L(t)$ scales linearly with
the system size gives an unbiased estimate of the critical temperature.
Eq.~\eqref{e.fss.At} implies that the scaling plot of
$A_L\,L^{-\rho/\nu}$ ~vs~ $y=(t-t_{\rm{c}}) L^{1/\nu}$, with a proper
estimate of $t_{\rm{c}}$, shows the data for different lattice sizes to
collapse onto the universal curve $F_A(y)$.

In the case of a BKT transition, in which no order parameter is given,
the presence of topological order at finite temperature is shown when
the helicity modulus scales to a finite value below a certain
temperature.  The use of the scaling Ansatz to locate the critical
temperature can be generalized to the case of a BKT transition, though
most of the critical exponents are not defined.  However the
Kosterlitz-Thouless theory predicts $\eta=1/4$ at the critical point,
so that a scaling behaviour of the susceptibility as
$L^{2-\eta}=L^{7/4}$ is a good signature of the critical temperature.
Moreover, Kosterlitz's renormalization group
equations~\cite{Kosterlitz74} provide the critical scaling law for the
helicity modulus in the form~\cite{OlssonM91}
\begin{equation}
\frac{\Upsilon_L(t_{_{\rm{BKT}}})}{t_{_{\rm{BKT}}}}
\approx \frac{2}{\pi}
\left( 1 + \frac{1}{2 \log(L/L_0)} \right)~,
\label{scalingYY}
\end{equation}
where $L_0$ is a constant. This relation has been widely used to locate
the BKT critical temperature of the classical 2D planar-rotator
model~\cite{WeberM88,Olsson95} and of the $S=1/2$  quantum XY
model~\cite{HaradaK98}.

We end this section with a general remark. It is observed that the
smaller the anisotropy, the bigger the lattice sizes required to enter
the asymptotic scaling regime, where FSS holds. This is essentially due
to the fact that the critical region is shifted to lower temperature:
the correlation length of the isotropic model, acting as a lower bound
for that of the nearly isotropic ones, increases exponentially upon
lowering the temperature, and therefore, keeping the lattice size
fixed, the ratio $L/\xi_L$, that drives the onset of asymptotic scaling
near the transition, gets smaller.

\section{Easy-axis model and Ising transition}
\label{s.EAXXZ}

The values of the anisotropy here considered are $\Delta_\mu=0.01$
(also used in Ref.~\onlinecite{QMC-EA}) and $\Delta_\mu=0.001$. They
are comparable with the characteristic anisotropies of real compounds;
yet, for such small anisotropy there is no universal consensus on the
existence of a transition~\cite{BrancoRdS00,RicardodeSousaA00}. From
previous works~\cite{QMC-EA,Aplesnin98,CRTVV01} the transition is
expected in the temperature range $0.2<t<0.3$ in both systems. At
higher temperature the behaviour gets closer to that of the isotropic
model, which has been extensively investigated by means of QMC in
recent years~\cite{MakivicD91,BeardW96,Beardetal98,KimT98}; we have
extended our analysis up to $t\simeq{0.8}$ in order to identify those
deviations from the isotropic behaviour that can be experimentally
detected above the critical region.

In our approach, evidences for the existence of an Ising-like
transition follow from a detailed FSS analysis of the data;
subsequently, we analyze the temperature dependence of some relevant
thermodynamic quantities, emphasizing the signatures of the EA nature.

\subsection{Finite-size scaling analysis}
\label{ss.EA-FSS}

Our analysis proceeds in three steps: we give evidence for a transition
to occur, then the transition temperature is located, and eventually
the Ising critical scaling is tested. After the discussion made in
Section~\ref{ss.FSS} the FSS analysis for $\Delta_\mu=0.001$ is
expected to be more delicate than for $\Delta_\mu=0.01$. Indeed, for
the lattice sizes used ($L\le{128}$) some quantities show to have well
entered the asymptotic scaling regime, while others do not. Anyway,
clear (though not complete) evidence for the Ising universality class
is given also for $\Delta_\mu=0.001$; larger lattices would be required
to reach a full characterization.

\begin{figure}
\includegraphics[bbllx=0mm,bblly=10mm,bburx=183mm,bbury=151mm,%
  width=80mm,angle=0]{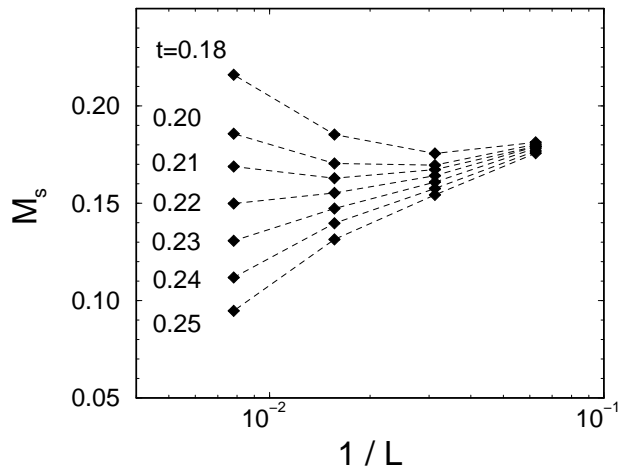}
 \caption{\label{EA-szscale}
Scaling of the staggered magnetization $M_{_{\rm{s}}}$ in the EA model
with $\Delta_\mu=0.001$, for different $t$.}
\end{figure}

Let us first consider the order parameter, i.e., the staggered
magnetization given in Eq.~\eqref{e.qmc.ms}. In Fig.~\ref{EA-szscale}
$M_{_{\rm{s}}}$ for $\Delta_\mu=0.001$ is seen to scale to a finite
value if $t\lesssim{0.22}$, so that the magnetization in the
thermodynamic limit becomes finite; the same behaviour is {\it a
fortiori} observed in the case $\Delta_\mu=0.01$.
We then invoke the scaling Ansatz~\eqref{e.fss.At} for the longitudinal
correlation length $\xi^{zz}$. The scaling plot of
$\xi_{\rm{fit}}^{zz}$, as specifically defined in
Section~\ref{ss.EA-xi} below, is shown in Fig.~\ref{EA-xiscale} for
$\Delta_\mu=0.001$ and gives
$t_{_{\rm{I}}}(\Delta_\mu{=}0.001)=0.2225(15)$. A similar analysis
yields $t_{_{\rm{I}}}(\Delta_\mu{=}0.01)=0.2815(25)$.

\begin{figure}
\includegraphics[bbllx=3mm,bblly=10mm,bburx=180mm,bbury=154mm,%
  width=80mm,angle=0]{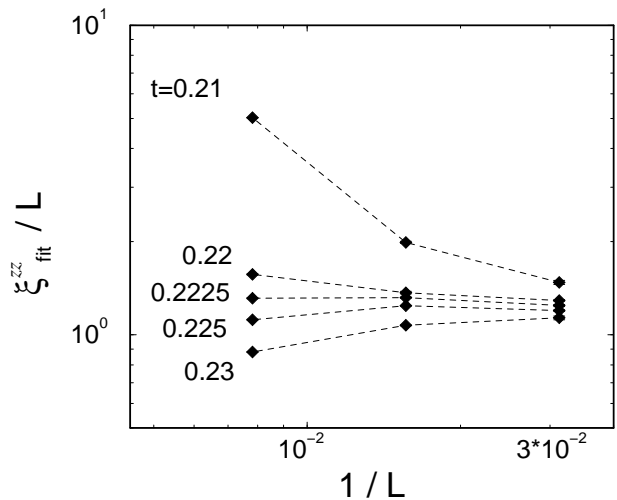}
 \caption{\label{EA-xiscale}
Scaling of the longitudinal correlation length $\xi_{{\rm{fit}}}^{zz}$
in the EA model with $\Delta_\mu=0.001$, for different $t$.}
\end{figure}

\begin{figure}
\includegraphics[bbllx=3mm,bblly=10mm,bburx=180mm,bbury=154mm,%
  width=80mm,angle=0]{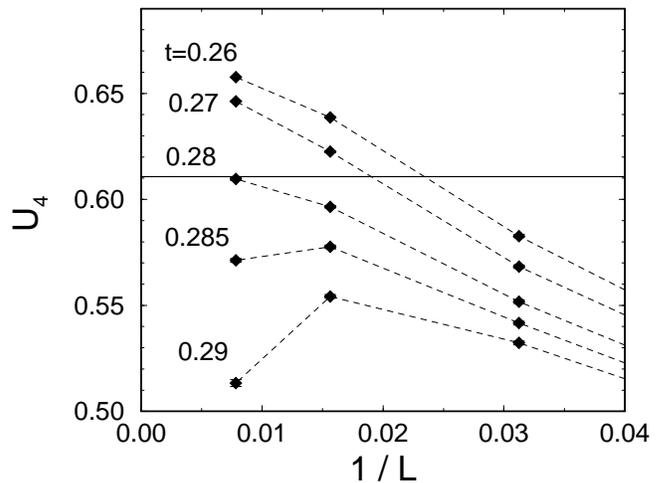}
 \caption{\label{EA-U4scale}
Scaling of the Binder's fourth cumulant in the EA model with
$\Delta_\mu=0.01$ for different $t$.  The solid line indicates the
universal critical value $U_4^{\rm(c)}$ (see text).}
\end{figure}

\begin{figure}
\includegraphics[bbllx=3mm,bblly=10mm,bburx=180mm,bbury=154mm,%
  width=80mm,angle=0]{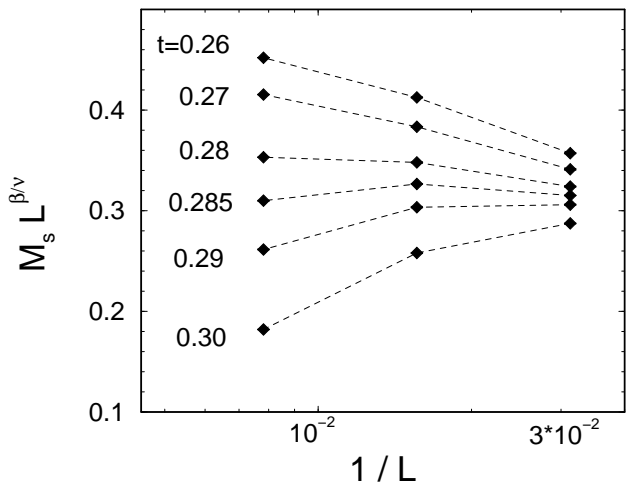}
 \caption{\label{EA-szscalemu.990}
Scaling of the staggered magnetization $M_{_{\rm{s}}}$ in the EA model
with $\Delta_\mu=0.01$, for different $t$.  The critical exponents
$\beta=1/8$ and $\nu=1$ are those of the Ising universality class.}
\end{figure}

Hitherto, no assumptions were made about the universality class. In
order to identify it, we consider the so called Binder's fourth
cumulant~\cite{Binder80}, shown in Fig.~\ref{EA-U4scale} and defined by
\begin{equation}
U_4 = 1 - \frac{\langle m_{_{\rm{s}}}^4 \rangle_{_{\rm{MC}}}} {3
\langle m_{_{\rm{s}}}^2 \rangle^2_{_{\rm{MC}}}}~,
\end{equation}
which assumes the universal critical value $U_4^{\rm(c)}=0.6106900(1)$
at $t_{_{\rm{I}}}$ in the 2D Ising model on the square
lattice~\cite{KamieniarzB93}, and increases (decreases) with $L$, below
(above) $t_{_{\rm{I}}}$. For $\Delta_\mu=0.01$, we verify such
behaviour and obtain $t_{_{\rm{I}}}(\Delta_\mu{=}0.01)=0.280(3)$,
consistently with the above unbiased estimate from the scaling of
$\xi^{zz}$.
The scaling Ansatz~\eqref{e.fss.At} for the staggered magnetization, $
M_{_{\rm{s}}}\sim{L^{-\beta/\nu}}$ at $t=t_{_{\rm{I}}}$, constitutes a
further way of checking the 2D Ising behaviour, since the critical
exponents $\beta=1/8$ and $\nu=1$ are involved. The data reported in
Fig.~\ref{EA-szscalemu.990} give $t_{_{\rm{I}}}=0.282(2)$.
In the case of the weakest anisotropy $\Delta_\mu=0.001$, both the
Binder's fourth cumulant and the staggered magnetization have not yet
well entered the asymptotic scaling region for the lattice sizes
considered, and $t_{_{\rm{I}}}$ cannot be reliably estimated by this
technique.

\begin{figure}
\includegraphics[bbllx=3mm,bblly=10mm,bburx=180mm,bbury=154mm,%
  width=80mm,angle=0]{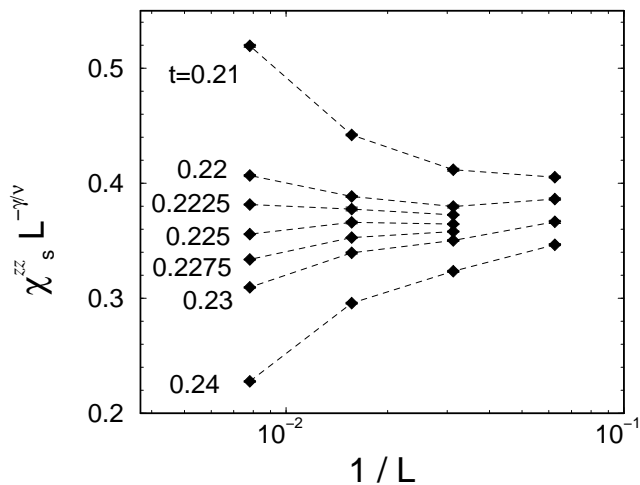}
 \caption{\label{EA-chiscalemu.999}
Scaling of the longitudinal staggered susceptibility
$\chi^{zz}_{_{\rm{s}}}$ in the EA model with $\Delta_\mu=0.001$, for
different $t$.  The critical exponents $\gamma=7/4$ and $\nu=1$ are
those of the Ising universality class.}
\end{figure}

A further test of the universality class involves the longitudinal
staggered susceptibility $\chi^{zz}_{_{\rm{s}}}$,
Eq.~\eqref{e.defs.chis}: in this case the scaling
Ansatz~\eqref{e.fss.At} gives
$\chi^{zz}_{_{{\rm{s}}L}}(t_{_{\rm{I}}})\sim{L^{\gamma/\nu}}$, with
2D-Ising critical exponents $\gamma=7/4$ and $\nu=1$, as shown in
Fig.~\ref{EA-chiscalemu.999} for the case $\Delta_\mu{=}0.001$. The
estimated critical temperatures result
$t_{_{\rm{I}}}(\Delta_\mu{=}0.01)=0.2825(25)$ and
$t_{_{\rm{I}}}(\Delta_\mu{=}0.001)=0.2235(15)$, in full agreement with
the above unbiased estimates.

To summarize, in the case $\Delta_\mu=0.01$ we find consistency for the
2D-Ising critical exponent ratios $\beta/\nu$ and $\gamma/\nu$, thus
fully verifying the universality class. For $\Delta_\mu=0.001$ the
evidence, though limited to the matching of the estimates of
$t_{_{\rm{I}}}$ obtained in Figs.~\ref{EA-xiscale}
and~\ref{EA-chiscalemu.999}, is quite convincing.

\begin{figure}
\includegraphics[bbllx=3mm,bblly=10mm,bburx=180mm,bbury=154mm,%
  width=80mm,angle=0]{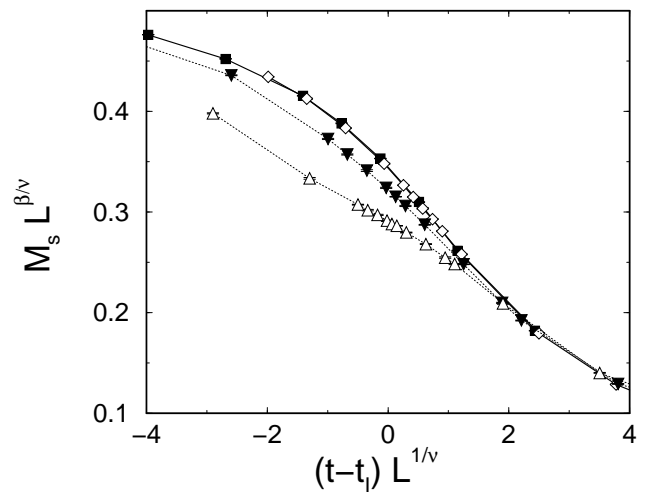}
 \caption{\label{EA-szscplotmu.990}
Scaling plot for the staggered magnetization $M_{_{\rm{s}}}$ in the EA
model with $\Delta_\mu=0.01$, for $L=16$ (up triangles), 32 (down
triangles), 64 (diamonds), 128 (squares). The critical exponents
$\beta=1/8$ and $\nu=1$ are those of the Ising universality class, and
the critical temperature is taken as $t_{_{\rm I}}=0.281$\,.}
\end{figure}

\begin{figure}
\includegraphics[bbllx=3mm,bblly=10mm,bburx=180mm,bbury=154mm,%
  width=80mm,angle=0]{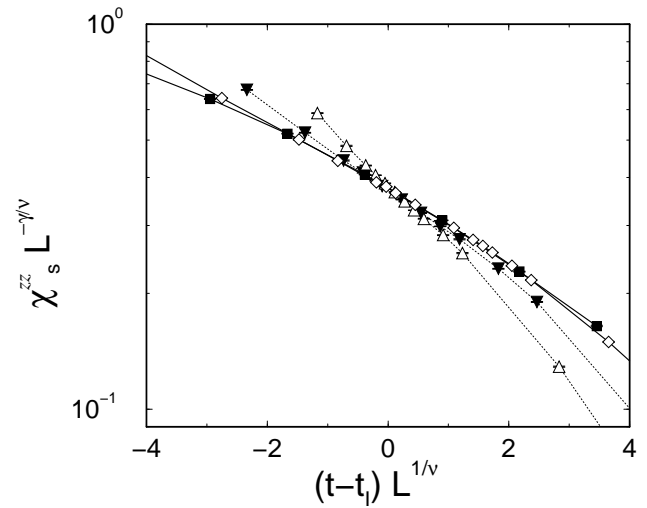}
 \caption{\label{EA-chiscplotmu.999}
Scaling plot for the longitudinal staggered susceptibility
$\chi^{zz}_{_{\rm{s}}}$ in the EA model with $\Delta_\mu=0.001$, for
different $L$; symbols as in Fig.~\ref{EA-szscplotmu.990}. The
universal scaling function emerges from the overlap of the two solid
lines. The critical exponents $\gamma=7/4$ and $\nu=1$ are those of the
Ising universality class, and the critical temperature is taken as
$t_{_{\rm I}}=0.223$\,.}
\end{figure}

As a check that the magnetization and the staggered susceptibility have
actually reached the asymptotic scaling regime with the considered
lattice sizes, we have constructed their scaling plots after
Eq.~\eqref{e.fss.At}, which are reported in
Figs.~\ref{EA-szscplotmu.990} and~\ref{EA-chiscplotmu.999}. Data
collapse for different lattice sizes is verified for the staggered
susceptibility in the case $\Delta_\mu=0.001$ for $L\geq64$, taking
$t_{_{\rm{I}}} = 0.223$, and {\it a fortiori} in the case
$\Delta_\mu=0.01$; the staggered magnetization is instead seen to have
entered the asymptotic scaling regime for $L\geq64$ only in the case
$\Delta_\mu=0.01$.

From the above analysis a strong indication for the existence of an
Ising phase transition is therefore given for both the considered
anisotropies. Estimates of the critical temperature
$t_{_{\rm{I}}}(\Delta_\mu)$ from the different criteria described in
this section are summarized in Table~\ref{t.ea.tc}; all estimates are
consistent, and amongst them we choose those realizing the best data
collapse onto the universal scaling function in the scaling plots of
staggered susceptibility and magnetization shown in
Fig.~\ref{EA-szscplotmu.990} and~\ref{EA-chiscplotmu.999}: the
resulting values are $t_{_{\rm{I}}}(0.01)=0.281(2)$ and
$t_{_{\rm{I}}}(0.001)=0.223(2)$.  Such values will be indicated with a
thin arrow in the following figures.

\begin{table}
\caption{\label{t.ea.tc}
2D Ising transition temperature $t_{_{\rm{I}}}(\Delta_\mu)$ as obtained
by FSS analysis and fit of critical behaviours.}
\begin{ruledtabular}
\begin{tabular}{lll}
 estimation method & \mbox{{$t_{_{\rm{I}}}(0.01)$}} &  \mbox{{$t_{_{\rm{I}}}(0.001)$}} \\
\colrule
 $\xi^{zz}\sim L$                        & 0.2815(25)  & 0.2225(15)  \\
 $U_4\to 0.6107$                         & 0.280(3)    &             \\
 $ M_{_{\rm{s}}}\sim L^{-\beta/\nu}$ & 0.282(2)    &             \\
 $\chi^{zz}_{_{\rm{s}}}\sim L^{\gamma/\nu}$          & 0.2825(25)  & 0.2235(15)  \\
\colrule
 $\xi^{zz}\sim |t-t_{_{\rm I}}|^{-\nu}$           & 0.283(6)    &           \\
 $\chi^{zz}_{_{\rm{s}}}\sim |t-t_{_{\rm I}}|^{-\gamma}$  & 0.284(4)   &      \\
\end{tabular}
\end{ruledtabular}
\label{t.Ising}
\end{table}

\subsection{Specific heat}
\label{ss.EA-spheat}

\begin{figure}
\includegraphics[bbllx=3mm,bblly=8mm,bburx=190mm,bbury=164mm,%
  width=80mm,angle=0]{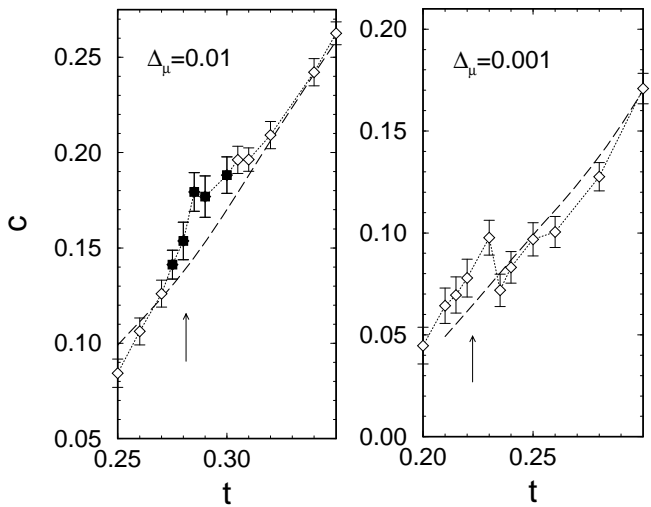}
 \caption{\label{EA-spheat}
Specific heat of the EA model vs $t$, for $L=64$ (diamonds), and $128$
(squares); the dashed line represent the specific heat of the isotropic
model, as obtained by numerically deriving the internal energy QMC data
of Ref.~\onlinecite{KimT98}. Arrows indicate the estimated critical
temperature.}
\end{figure}

The specific heat of the Ising model is characterized by a sharp peak
at the transition temperature. As the anisotropy decreases, a large
bump,  eventually coinciding with the bump of the isotropic model,
grows on the right-hand side of the peak, which correspondingly moves
towards lower temperatures, meanwhile getting narrower. In
Fig.~\ref{EA-spheat} we see that traces of an Ising-like peak emerging
from the isotropic curve can still be evidenced for both anisotropy
values. Despite their being traces, we observe that they develop at the
critical temperature as estimated above. These findings are in good
qualitative agreement with the experimental data~\cite{Algraetal78}
relative to the layered $S=1/2$ antiferromagnet
Cu(C$_5$H$_5$NO)$_6$(BF$_4$)$_2$, which is supposed to have an
anisotropy-driven transition; similar behaviour is displayed by
larger-spin compounds whose anisotropy is known to be
Ising-like~\cite{spheat-largeS}, such as K$_2$NiF$_4$ ($S=1$) and
K$_2$MnF$_4$ ($S=5/2$).

\subsection{Uniform susceptibility}
\label{ss.EA-chi0}

\begin{figure}
\includegraphics[bbllx=3mm,bblly=10mm,bburx=180mm,bbury=154mm,%
  width=80mm,angle=0]{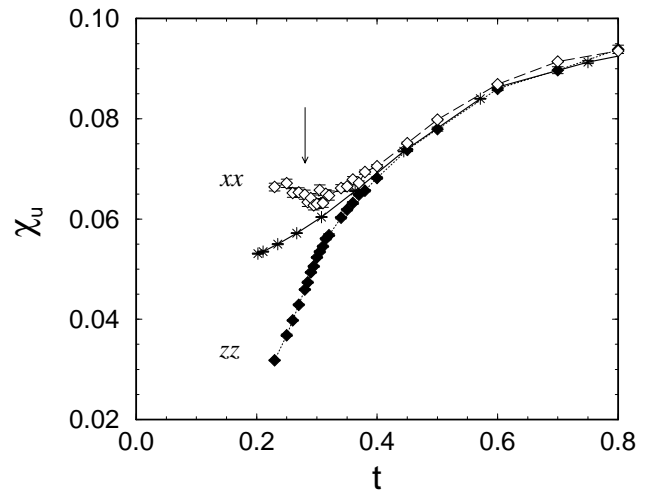}
 \caption{\label{EA-chi0}
Uniform susceptibility of the EA model for $\Delta_\mu=0.01$ and
$L=64$. Full diamonds: longitudinal; open diamonds: transverse branch;
stars: QMC data for the isotropic model~\cite{KimT98}. Solid and dashed
lines are guides to the eye. The arrow indicates the estimated critical
temperature.}
\end{figure}

At variance with the specific heat, where the anisotropic curves just
slightly differ from the isotropic one, the uniform susceptibility
undoubtedly shows an anisotropic behaviour: in Fig.~\ref{EA-chi0},
where data relative to $\Delta_\mu=0.01$ are shown, the transverse and
longitudinal components, $\chi_{_{_{\rm u}}}^{xx}$ and
$\chi_{_{_{\rm{u}}}}^{zz}$, separate from the isotropic curve at
$t\lesssim0.4$, i.e., well above $t_{_{\rm{I}}}=0.282$. It is quite
surprising that the Hamiltonian symmetry puts up so much resistance to
the disordering effects of both quantum and thermal fluctuations: this
means that the anisotropy, even as weak as those we are here
considering, can never be neglected, and that there exists a
temperature range, extending well above the transition (i.e., also out
of the region where 2D correlations can trigger the onset of 3D
long-range order), where genuinely 2D anisotropic behaviour can be
experimentally observed.

The different temperature dependence of the transverse and longitudinal
branches, with the former displaying a minimum and the latter
monotonically going to zero, is that expected for an EA
antiferromagnet. This behaviour results from the anisotropy-induced
spin ordering, that makes the system more sensitive to the application
of a transverse magnetic field, rather than of a longitudinal one.  We
observe that both the minimum of the in-plane component and the start
of the rapid decrease of the longitudinal one, are close to the
transition: as such feature is peculiar to the Ising model, this result
gives further strength to the characterization of the transition as of
Ising type.

The two components of the uniform susceptibility are experimentally
observable by means of conventional magnetometry measurements: the
above discussed deviations from the isotropic behaviour have been
actually observed in several layered compounds with $S\geq1$:
K$_2$NiF$_4$~\cite{Matsuuraetal70},
Rb$_2$NiF$_4$~\cite{Matsuuraetal70}, BaNiF$_4$~\cite{Coxetal70}
($S=1$), K$_2$MnF$_4$~\cite{Breed66-67},
Rb$_2$MnF$_4$~\cite{Breed66-67}, BaMnF$_4$~\cite{Holmesetal69}
($S=5/2$). Such effects are here proved to be substantial also in
$S=1/2$ systems with a comparable anisotropy; unfortunately, to our
knowledge, no clean experimental realization of a 2D $S=1/2$ HAFM with
small EA anisotropy is available yet.

\subsection{Staggered susceptibility}
\label{ss.EA-chist}

\begin{figure}
\includegraphics[bbllx=3mm,bblly=10mm,bburx=180mm,bbury=154mm,%
  width=80mm,angle=0]{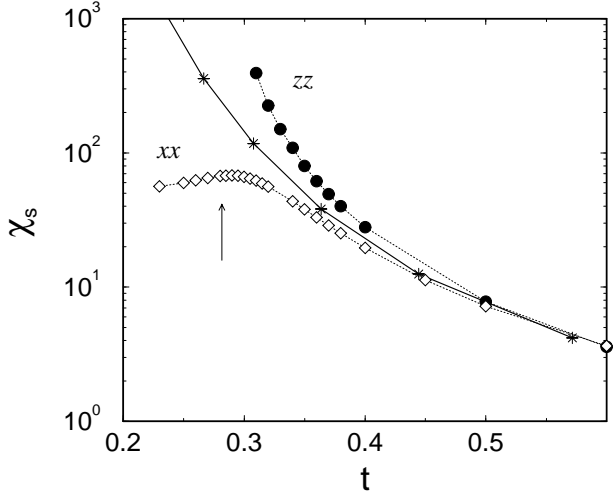}
 \caption{\label{EA-chistagg}
Staggered susceptibility of the EA model for $\Delta_\mu=0.01$.
Circles: longitudinal (bulk values); diamonds: transverse ($L=64$);
stars, lines and arrow as in Fig.~\ref{EA-chi0}.}
\end{figure}

The equal-time longitudinal and transverse staggered susceptibilities,
$\chi^{zz}_{_{\rm{s}}}$ and $\chi^{xx}_{_{\rm{s}}}$, for
$\Delta_\mu=0.01$ are shown in Fig.~\ref{EA-chistagg}, together with
the susceptibility of the isotropic model~\cite{KimT98}. Below the
high-temperature region where the isotropic behaviour is reproduced,
the two curves separate at $t\simeq0.4$, below which
$\chi^{zz}_{_{\rm{s}}}$ diverges more rapidly than in the isotropic
case, while $\chi^{xx}_{_{\rm{s}}}$ stays finite and shows a maximum at
about the transition temperature. The time-averaged susceptibilities
display the same qualitative behaviour, though their values are
slightly different with respect to the equal-time case.

As in the case of the uniform susceptibility, the observed behaviour is
qualitatively suggestive of an Ising-like transition.  Moreover, the
analysis of longitudinal branch divergence gives a direct evidence of
the Ising universality class, as well as an independent estimate of the
critical temperature. For a 2D-Ising transition $\chi^{zz}_{_{\rm{s}}}$
must display a power-law divergence,
$\chi^{zz}_{_{\rm{s}}}\sim|t-t_{_{\rm{I}}}|^{-\gamma}$, with
$\gamma=7/4$.  In Fig.~\ref{EA-powerlaw} we plot
$(\chi^{zz}_{_{\rm{s}}})^{-1/\gamma}$ vs $t$ for $\Delta_\mu=0.01$,
using data which are free of significant finite-size corrections,
according to the criteria described in Section~\ref{ss.EA-xi}.  The
power-law with the Ising exponent $\gamma=7/4$ is evidently verified
and the extrapolated critical temperature is $t_{_{\rm{I}}}=0.284(4)$,
which agrees with the more accurate value obtained in
Section~\ref{ss.EA-FSS}. As for the smaller anisotropy,
$\Delta_\mu=0.001$, the power-law divergence of $\chi_{_{\rm{s}}}^{zz}$
could not be unambiguously detected for the considered lattice sizes.

\subsection{Correlation length}
\label{ss.EA-xi}

\begin{figure}
\includegraphics[bbllx=3mm,bblly=10mm,bburx=180mm,bbury=154mm,%
  width=80mm,angle=0]{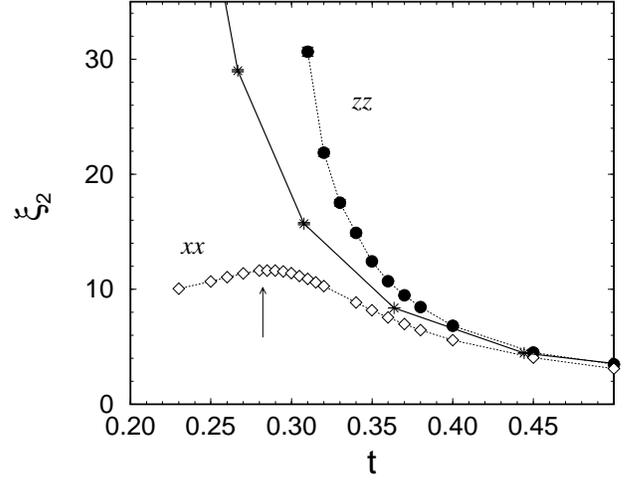}
 \caption{\label{EA-xi}
Correlation length of the EA model for $\Delta_\mu=0.01$. Symbols,
lines and arrow as in Fig.~\ref{EA-chistagg}.}
\end{figure}

Fig.~\ref{EA-xi} shows the longitudinal and the transverse correlation
lengths, $\xi^{zz}$ and $\xi^{xx}$, for $\Delta_\mu=0.01$. The two
correlation lengths behave quite differently: the transverse branch,
after having left the longitudinal one at a temperature $t\simeq{0.4}$,
displays a maximum at the transition, while the longitudinal branch
diverges faster than in the isotropic model. Again, the overall
behaviour is suggestive of a 2D Ising transition.

\begin{figure}
\includegraphics[bbllx=3mm,bblly=10mm,bburx=180mm,bbury=154mm,%
  width=80mm,angle=0]{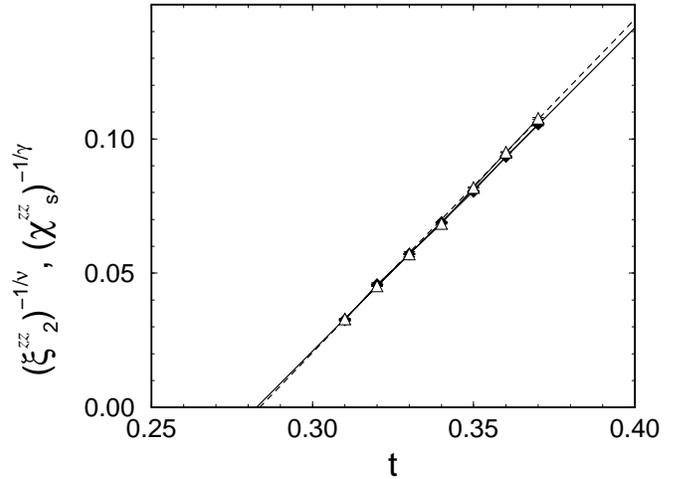}
 \caption{\label{EA-powerlaw}
Power-law critical behaviour of the longitudinal correlation length
$\xi^{zz}$ (diamonds) and of the longitudinal staggered susceptibility
$\chi^{zz}_{_{\rm{s}}}$ (open triangles) for $\Delta_\mu=0.01$; solid
and dashed line are linear fits of $(\xi^{zz})^{-1/\nu}$ and
$(\chi^{zz}_{_{\rm{s}}})^{-1/\gamma}$ respectively. The critical
exponents $\nu=1$ and $\gamma=7/4$ are those of the Ising universality
class.}
\end{figure}

The longitudinal antiferromagnetic correlation length $\xi^{zz}$ is
expected to display a power-law divergence
$\xi^{zz}\sim|t-t_{_{\rm{I}}}|^{-\nu}$, with $\nu=1$. One can capture
this divergence by selecting a few points for $\xi_{2}^{zz}$ at
temperatures immediately above $t_{_{\rm{I}}}$, discarding those
exceeding $L/4$, which are affected by finite-size saturation. This
criterion is reinforced by requiring the consistency of the estimates
of $\xi_{2}^{zz}$ obtained via the equal-time- and the time-averaged
susceptibilities: since both estimates converge to the same value in
the thermodynamic limit, their agreement indicates that finite-size
effects are under control. For $\Delta_\mu=0.01$,
Fig.~\ref{EA-powerlaw} shows that $(\xi^{zz})^{-1}$ is linear, with an
extrapolated intercept $t_{_{\rm{I}}}=0.283(6)$, in agreement with the
value found via FSS analysis.

The same observation is not possible for $\Delta_\mu=0.001$, as sizes
larger than those here considered are required to approach the critical
point of such model, while controlling finite-size effects.

We have also extracted the longitudinal correlation length $\xi^{zz}$
in the vicinity of the critical point by fitting the equal-time
correlator $C^{zz}({\bm r})$, defined in Eq.~\eqref{e.defs.cr}, to a
function due to Serena, Garc\'\i a and Levanyuk~\cite{Serenaetal93},
\begin{equation}
 F(x) = \frac{e^{-x}}{x^{1/2}+x^{1/4}}~,
\end{equation}
properly symmetrized so as to take into account the periodic boundary
conditions, i.e., by
\begin{equation}
 C^{zz}(r) \propto F\big(r/\xi^{zz} \big)
  + F\big((L-r)/\xi^{zz} \big)~.\label{symmcorr}
\end{equation}
This function interpolates between the known asymptotic behaviours at
$r\to0$ and $r\to\infty$ of the Ising model. Well above the critical
point we used the conventional fitting function for the isotropic
antiferromagnet~\cite{MakivicD91}
\begin{equation}
 F(x)=\frac{e^{-x}}{x^\eta}~.
\label{isocorr}
\end{equation}
In the case $\Delta_\mu=0.001$ good and stable fits are obtained even
if the correlation length becomes comparable to (or even exceeds) the
lattice size $L$: we can hence univocally define the fitted correlation
length $\xi^{zz}\equiv\xi_{\rm fit}^{zz}$.  Moreover, the same kind of
fitting procedure on the time-averaged correlator $C^{zz}(r)$,
Eq.~\eqref{e.defs.cr}, gives consistent results.

\begin{figure}
\includegraphics[bbllx=3mm,bblly=10mm,bburx=180mm,bbury=154mm,%
  width=80mm,angle=0]{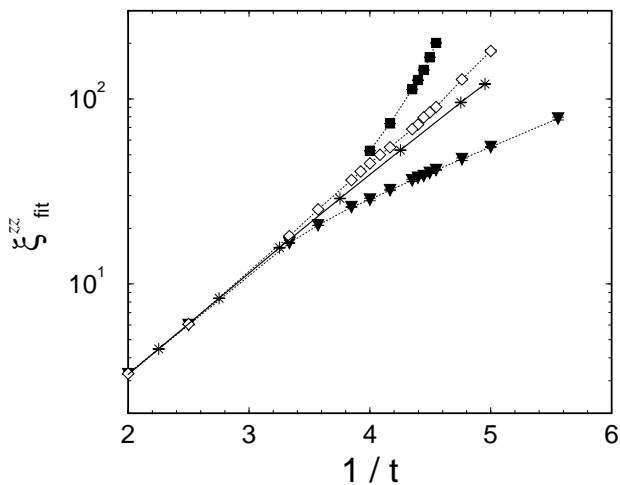}
 \caption{\label{EA-ximu.999}
Longitudinal correlation length $\xi_{\rm fit}^{zz}$ of the EA model
with $\Delta_\mu=0.001$, for $L=32$ (down triangles), 64 (diamonds),
128 (squares). Stars and lines as in Fig.~\ref{EA-chi0}.}
\end{figure}

Notice that $\xi_{{\rm fit}}^{zz}$ monotonically increases with $L$,
bounded from above by the thermodynamic value; on the other hand, as
$\xi_{2}^{zz}$ is systematically smaller than $\xi_{{\rm fit}}^{zz}$,
the latter is necessarily less sensitive to size finiteness. For this
reason it is possible to observe in Fig.~\ref{EA-ximu.999} the clear
deviation of $\xi_{{\rm fit}}^{zz}$ from the isotropic model, due to
its divergence at $t_{_{\rm{I}}}$.
To summarize, the sharp dependence of the longitudinal correlation
length to small anisotropies, already observed for $S=5/2$ in
Rb$_2$MnF$_4$~\cite{Leeetal98} and KFeF$_4$~\cite{Fultonetal94}, is
evidenced also for $S=1/2$.

\section{Easy-plane model and BKT transition}
\label{s.EPXXZ}

In this section we present our results relative to the EP model.  We
have used lattice sizes up to $L=200$ and two anisotropy values:
$\Delta_\lambda=0.02$ (already considered in
Ref.~\onlinecite{Ding92-EP}) and $\Delta_\lambda=0.001$.  These values
are comparable with the experimentally estimated anisotropies of real
compounds, amongst which several cupreous oxides such as La$_2$CuO$_4$,
Sr$_2$CuO$_2$Cl$_2$, and Pr$_2$CuO$_4$, which are known to have an EP
anisotropy~\cite{Johnston97}.

The temperature range covered by our simulations is
$0.15\lesssim{t}\lesssim 0.90$~: as suggested by previous
calculations~\cite{Ding92-EP,CRTVV01} this is the interval where we
expect most of the peculiar features due to the anisotropy to be
detectable. At higher temperatures the thermodynamic behaviour of the
model does not differ from that of the isotropic one. On the other
hand, finite-size limitations preclude the study of the very-low
temperature region. To this respect, we recall that the correlation
length of an EP model is expected to diverge exponentially as
$t\to{}t_{_{\rm{BKT}}}^+$; such fast divergence makes finite-size
limitations more severe than in the EA case, where $\xi$ diverges
algebraically. On the whole, the BKT transition offers less robust
evidences, both numerically and experimentally, due to its being a
topological phase transition rather than a second-order one.

In what follows we will refer to {\it out-of-plane} quantities as those
related to the hard $z$-axis, and to {\it in-plane} quantities as those
related to the easy $xy$-plane.

\subsection{Finite-size scaling analysis}

\begin{figure}

\includegraphics[bbllx=0mm,bblly=10mm,bburx=180mm,bbury=151mm,%
  width=80mm,angle=0]{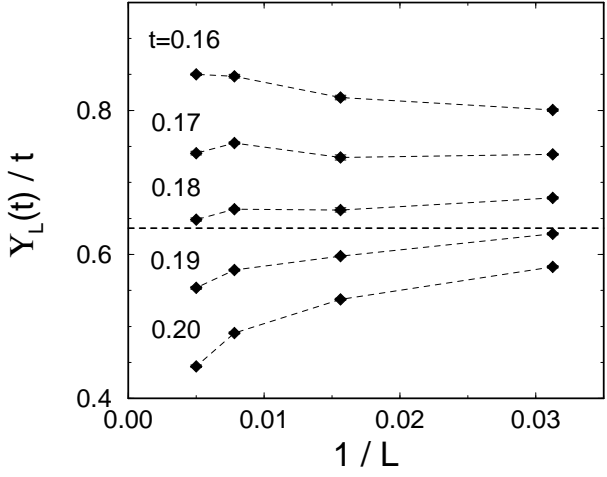}
 \caption{\label{EP-YYonTscalela.999}
Scaling of $\Upsilon/t$ in the EP model for $\Delta_\lambda=0.001$. The
horizontal dashed line indicates the value $2/\pi$.}
\end{figure}

The role of the staggered magnetization in the FSS analysis of the EA
behaviour is somehow taken, in the EP case, by the helicity modulus
$\Upsilon$, defined in Sec.~\ref{ss.defs}. In the thermodynamic limit
$\Upsilon$ is finite below and vanishes above the transition. When
finite-size systems are considered, the occurrence of a BKT transition
is marked by the existence of a finite temperature below which
$\Upsilon$ scales to a finite value, as suggested by
Fig.~\ref{EP-YYonTscalela.999} for $\Delta_\lambda=0.001$.

\begin{figure}
\null\hspace{-1cm}
\includegraphics[bbllx=1mm,bblly=9mm,bburx=197mm,bbury=155mm,%
  width=80mm,angle=0]{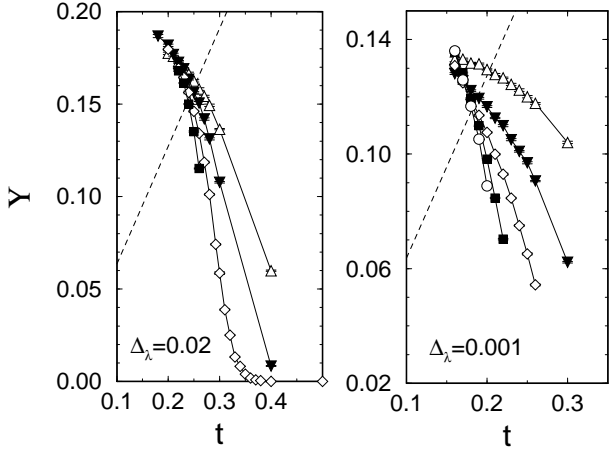}
 \caption{\label{EP-YY}
Helicity modulus of the EP model for $L=16$ (up triangles), 32 (down
triangles), 64 (diamonds), 128 (squares), 200 (circles). The dashed
line is the function $2t/\pi$.}
\end{figure}
As for the value of the critical temperature, one knows that in the
thermodynamic limit the ratio $\Upsilon/t$ gets the universal value
$2/\pi$ at the transition~\cite{NelsonK77}. This behaviour is clearly
detected in Fig.~\ref{EP-YY}, where the helicity modulus is shown vs
temperature for different $L$: the slope of $\Upsilon(t)$ near the
point where the line $2t/\pi$ is crossed, gets larger for larger sizes,
consistently with the occurrence of a jump in the thermodynamic limit.

An upper bound to the BKT critical temperature can be hence given by
looking at the temperature $t$ where the scaling behaviour of
$\Upsilon/t$ is most compatible with the expected asymptotic value
$2/\pi$ at criticality. From Fig.~\ref{EP-YYonTscalela.999} we obtain
$t_{_{\rm{BKT}}}(\Delta_\lambda{=}0.001)\lesssim0.180$; the same
procedure leads to
$t_{_{\rm{BKT}}}(\Delta_\lambda{=}0.02)\lesssim0.235$.

\begin{figure}
\includegraphics[bbllx=-5mm,bblly=10mm,bburx=199mm,bbury=157mm,%
  width=80mm,angle=0]{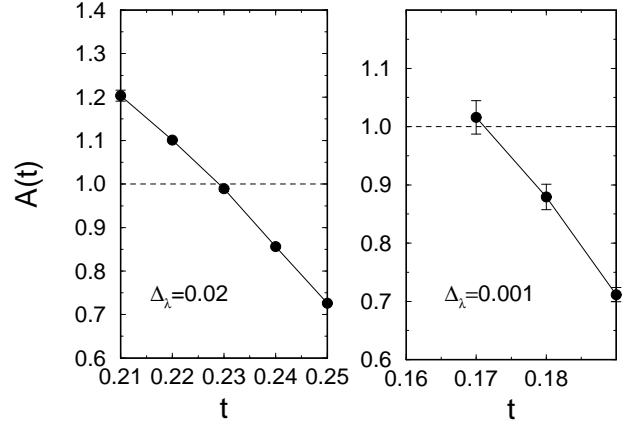}
 \caption{\label{EP-AvsT}
Fitting parameter $A$ vs $t$. The crossing point with the line $A=1$
gives an estimate of the critical temperature $t_{_{\rm{BKT}}}$.}
\end{figure}

More accurate results are obtained by considering the critical scaling
of $\Upsilon$, Eq.~\eqref{scalingYY}. According to the procedure
suggested in Ref.~\onlinecite{HaradaK98}, we assume the relation
\begin{equation}
\frac{\Upsilon_L(t)}{t}=
\frac{2A(t)}{\pi}\left(1+\frac{1}{2\log(L/L_0)}\right)
\end{equation}
to hold in the vicinity of the transition; $A(t)$ and $L_0$ are then
determined {\it via} a best-fit procedure and $t_{_{\rm{BKT}}}$
identified as the temperature where $A(t)$ equals unity, as shown in
Fig.~\ref{EP-AvsT}. The resulting estimates are
$t_{_{\rm{BKT}}}(\Delta_\lambda{=}0.02)=0.229(2)$ and
$t_{_{\rm{BKT}}}(\Delta_\lambda{=}0.001)=0.172(5)$. In the case
$\Delta_\lambda=0.001$, this procedure is more uncertain: due to strong
finite-size effects, $\Upsilon$ is seen to asymptotically scale just
for $L\gtrsim128$ (to be compared with $L\gtrsim32$ in the case
$\Delta_\lambda=0.02$), so that the logarithmic fit can only be
performed on two points ($L=128,~200$) for each temperature.

\begin{figure}
\includegraphics[bbllx=3mm,bblly=10mm,bburx=180mm,bbury=154mm,%
  width=80mm,angle=0]{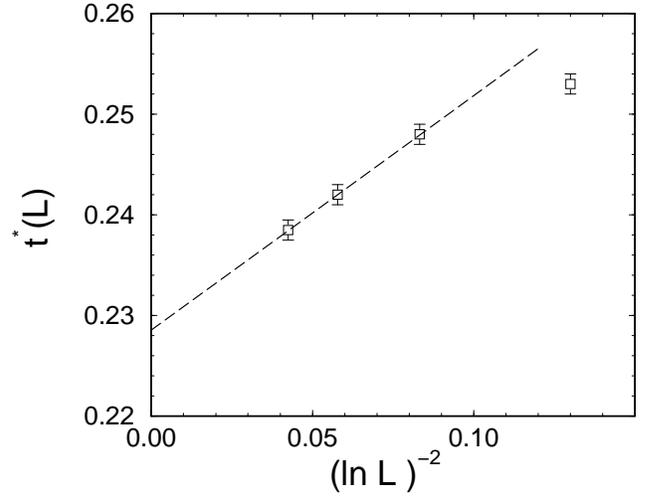}
 \caption{\label{EP-Tstarscaling}
Scaling of $t^*$ with $L$ for $\Delta_\lambda=0.02$. The dashed line is
a linear fit of the first three points, corresponding to $L=32$, 64,
128.}
\end{figure}

There is another way to exploit the data for the helicity modulus of a
finite-size system, though we are not aware of such technique having
been used by other authors before. In
Ref.~\onlinecite{BramwellH93jpcm}, Bramwell and Holdsworth found that
in the classical 2D planar-rotator model on a finite size the ratio
$\Upsilon/t$ gets the universal value $2/\pi$ at a temperature
$t^*>t_{_{\rm{BKT}}}$, whose size dependence is given by
\begin{equation}
t^*\simeq t_{_{\rm{BKT}}}+\frac{\pi^2}{4c(\ln L)^2}~.
\label{e.t*BH}
\end{equation}
The above relation, determined by a renormalization-group based
approach, is presented as a general property of BKT systems, though to
our knowledge its validity has never been checked for models others
than the classical pure planar one.  On the other hand, in the case
$\Delta_\lambda=0.02$ we can easily determine $t^*$ as a function of
$L$ from Fig.~\ref{EP-YY} and hence get Fig.~\ref{EP-Tstarscaling},
which shows that Eq.~\eqref{e.t*BH} holds even for weakly anisotropic,
strongly quantum models; a linear fit of the scaling behaviour of $t^*$
against $(\ln L)^{-2}$ for $L\geq32$ provides us with a rather accurate
estimate of the critical temperature,
$t_{_{\rm{BKT}}}(\Delta_\lambda{=}0.02)=0.228(4)$.  Moreover, the
results of Ref.~\onlinecite{BramwellH93jpcm} relate the coefficient $c$
to the coefficient $b_{\xi}$ appearing in Eq.~\eqref{chixicritical} in
the form $b_{\xi}=\pi/\sqrt{c}$, and from the linear fit we obtain
$b_{\xi} = 0.96(9)$, in good agreement with the value obtained below by
fitting the critical behaviour of the correlation length. This
remarkably shows that the predictions of
Ref.~\onlinecite{BramwellH93jpcm}, derived for the classical 2D
planar-rotator model, fully apply also to the quantum nearly-isotropic
antiferromagnet we considered.

\begin{figure}
\includegraphics[bbllx=3mm,bblly=10mm,bburx=180mm,bbury=154mm,%
  width=80mm,angle=0]{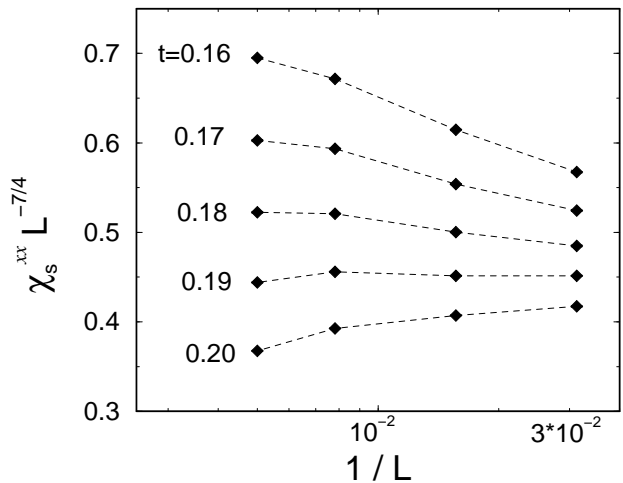}
 \caption{\label{EP-chixyscalela.999}
Scaling of the in-plane staggered susceptibility $\chi^{xx}_s$ in the
EP model with $\Delta_\lambda=0.001$.}
 \end{figure}

Finally, an additional estimate of the BKT critical temperature is
obtained by the in-plane staggered susceptibility, which is expected to
scale at the transition as $L^{2-\eta}$ with $\eta=1/4$. Looking for
the temperature where this scaling law is best verified, we obtain
$t_{_{\rm{BKT}}}(\Delta_\lambda{=}0.02)=0.230(5)$ and
$t_{_{\rm{BKT}}}(\Delta_\lambda{=}0.001)=0.180(5)$, as shown in
Fig.~\ref{EP-chixyscalela.999} for the latter value.

\begin{table}
\caption{\label{t.ep.tc}
BKT transition temperatures $t_{_{\rm{BKT}}}(\Delta_\lambda)$ as
obtained by FSS analysis and fit of critical behaviour of $\xi^{xx}$.}
\begin{ruledtabular}
\begin{tabular}{lll}
 estimation method & \mbox{{$t_{_{\rm{BKT}}}(0.02)$}} &
 \mbox{{$t_{_{\rm{BKT}}}(0.001)$}} \\
\colrule
 asymptotic value of $\Upsilon$      & $\lesssim$ 0.235 & $\lesssim$ 0.175 \\
 $A(t) = 1$                            & 0.229(2)    &   0.172(5)  \\
 scaling of $t^*(L)$                  & 0.228(4)    & \\
 $\chi^{xx}_{_{\rm{s}}}\sim L^{2-\eta}$          & 0.230(5)  & 0.180(5)  \\
%\colrule
 $\xi^{xx}\sim \exp[~b_{\xi}(t{-}t_{_{\rm{BKT}}})^{-1/2}] $  & 0.235(6)   &  \\
\end{tabular}
\end{ruledtabular}
\label{t.tbkteta}
\end{table}

Although the identification of the BKT universality class is less
complete than in the Ising case, substantial consistence between the
different estimates of the critical temperature obtained by different
predictions of the Kosterlitz-Thouless theory proves that the two
anisotropic models display a BKT critical regime. The estimates for the
critical temperature $t_{_{\rm{BKT}}}(\Delta_\lambda)$ given in this
section are summarized in Table~\ref{t.ep.tc} for the two anisotropies
considered. Putting together these estimates we choose as reference
values $t_{_{\rm{BKT}}}(\Delta_\lambda{=}0.02)=0.229(5)$ and
$t_{_{\rm{BKT}}}(\Delta_\lambda{=}0.001)=0.175(10)$. Such values will
be indicated by a thin arrow in the figures of the following sections.

\subsection{Specific heat}

\begin{figure}
\includegraphics[bbllx=3mm,bblly=10mm,bburx=180mm,bbury=154mm,%
  width=80mm,angle=0]{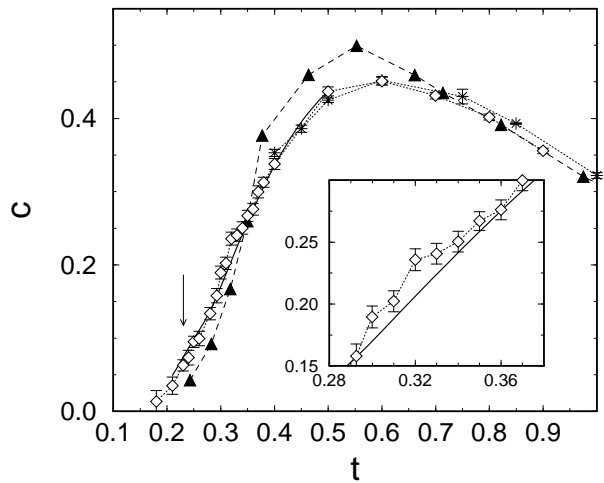}
 \caption{\label{EP-cv}
Specific heat of the EP model with $\Delta_\lambda=0.02$ for $L=64$
(diamonds) compared to QMC data for the same model~\cite{Ding92-EP}
(triangles) and for the isotropic model~\cite{MakivicD91,KimT98}
(stars). The arrow indicates the estimated BKT temperature. Inset: zoom
on the temperature region where a deviation is observed with respect to
the isotropic case.}
\end{figure}

The specific heat does not show large systematic deviations from the
isotropic case within the resolution reached by the simulations for
both anisotropies considered. Only a small temperature region, well
above the estimated transition temperature, displays an anomaly in the
form of a tiny peak, as shown in Fig.~\ref{EP-cv}; such peak is
possibly reminiscent of the rounded peak shown by the specific heat of
the quantum $S=1/2$ XY model above its BKT transition~\cite{Ding92-XY}.
We must however mention that, at variance with our results, previous
QMC data~\cite{Ding92-EP}, also reported in Fig.~\ref{EP-cv}
significantly deviate from the isotropic model. According to the
generally low sensitivity shown by the specific heat to weak
anisotropies, as seen for instance in the EA case, we find this result
a bit unlikely.

\subsection{Uniform susceptibility}

\begin{figure}
\includegraphics[bbllx=3mm,bblly=10mm,bburx=180mm,bbury=154mm,%
  width=80mm,angle=0]{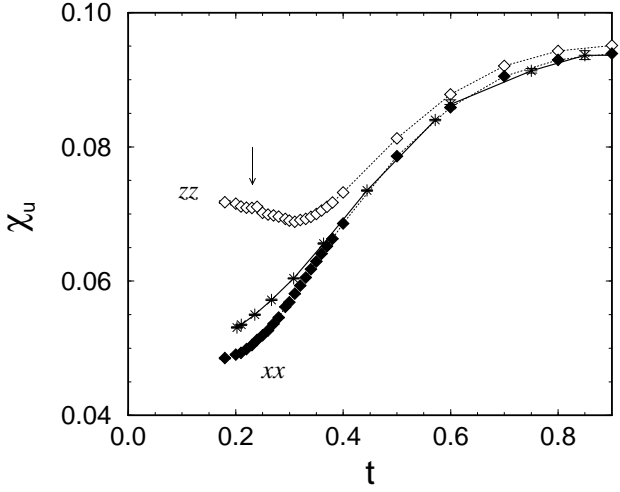}
 \caption{\label{EP-chi0}
Uniform susceptibility of the EP model with $\Delta_\lambda=0.02$ and
$L=64$. Full diamonds: in-plane; open diamonds: out-of-plane; stars:
QMC data for the isotropic model~\cite{KimT98,MakivicD91}. Solid and
dashed lines are guides to the eye. The arrow indicates the estimated
BKT temperature.}
\end{figure}

As in the EA case, the uniform susceptibility reported in
Fig.~\ref{EP-chi0} shows strong evidences of the anisotropy. Moving
down from the high-temperature region, where the isotropic behaviour is
reproduced, the in-plane $\chi^{xx}_{_{\rm u}}$ and out-of-plane
$\chi^{zz}_{_{\rm u}}$ uniform susceptibilities separate from each
other and from the isotropic data.

The in-plane component decreases faster than in the isotropic case. At
variance with the EA case, however, $\chi^{xx}_{_{\rm u}}$ is not
expected to vanish at $t=0$, due to the continuous rotational symmetry
of the ground state in the $xy$ plane. Indeed, in a semiclassical
picture, such symmetry allows the staggered magnetization to align
along the in-plane axis perpendicular to the field, making possible the
canted spin configuration with a finite ferromagnetic magnetization
parallel to -and linear in- the field, so that $\chi^{xx}_{_{\rm u}}$
stays finite.

The out-of-plane susceptibility is instead enhanced with respect to the
isotropic case, showing a minimum well above the transition. Such
minimum marks the onset of a completely different behaviour with
respect to the isotropic model, entirely due to the presence of the
small anisotropy.

\subsection{Staggered susceptibility and correlation length}

\begin{figure}
\includegraphics[bbllx=3mm,bblly=10mm,bburx=180mm,bbury=154mm,%
  width=80mm,angle=0]{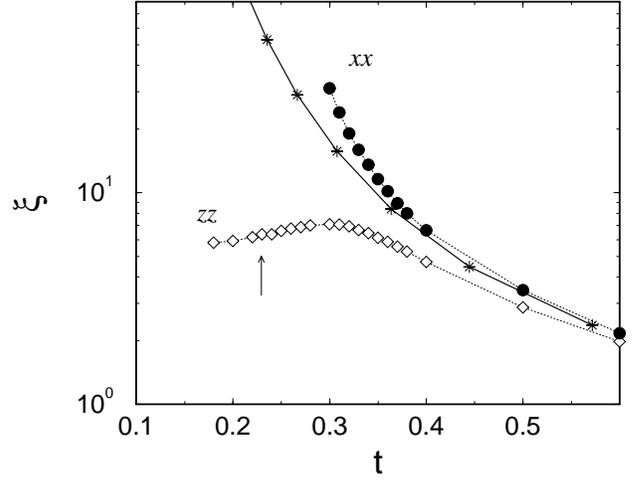}
 \caption{\label{EP-xi} Correlation length of the EP model with
$\Delta_\lambda=0.02$. Circles: in-plane (bulk values); diamonds: out
of plane ($L=64$); stars, solid line and arrow as
in Fig.~\ref{EP-chi0}.}
\end{figure}

\begin{figure}
\includegraphics[bbllx=3mm,bblly=10mm,bburx=180mm,bbury=154mm,%
  width=80mm,angle=0]{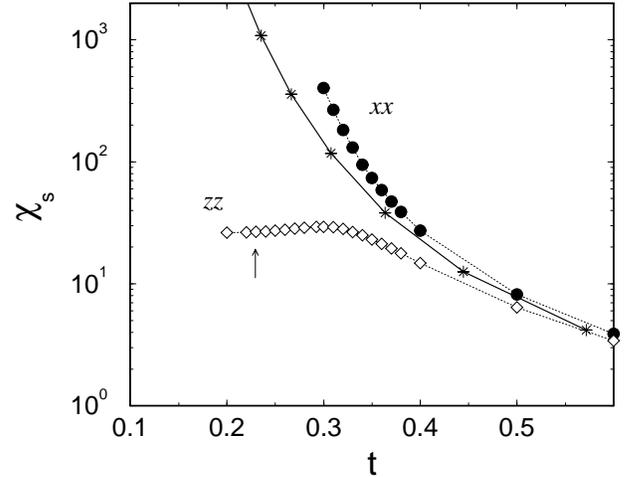}
 \caption{\label{EP-chi}
Staggered susceptibility of the EP model with $\Delta_\lambda=0.02$.
Symbols as in Fig.~\ref{EP-xi}; stars, lines and arrow as in
Fig.~\ref{EP-chi0}.}
\end{figure}

According to Kosterlitz theory~\cite{Kosterlitz74}, in presence of a
BKT transition the correlation length $\xi^{xx}$ is expected to diverge
exponentially at finite temperature as
\begin{equation}
 \xi^{xx} = a_{\xi} ~\exp\big[b_\xi(t-t_{_{\rm{BKT}}})^{-1/2}\big]~.
\label{chixicritical}
\end{equation}
As for the $\Delta_\lambda=0.02$ model, this behaviour is in fact
observed in Ref.~\onlinecite{Ding92-EP}, where it is used to estimate
the critical temperature. We use the estimates $\xi^{xx}_{_{\rm{fit}}}$
obtained by fitting the in-plane correlation function to
Eqs.~\eqref{symmcorr} and~\eqref{isocorr}. Discarding the values
affected by finite-size saturation and thus considering only those
satisfying $\xi^{xx}\lesssim{L/4}$, we also observe the predicted
behaviour: in particular, singling out the BKT critical region by
successively dropping points at high temperature until a stable fit is
obtained, we obtain $a_\xi=0.6(2)$, $b_\xi=1.0(1)$, and the estimate
$t_{_{\rm{BKT}}}=0.235(6)$, which agrees with the value found via FSS
analysis.

\begin{figure}
\includegraphics[bbllx=0mm,bblly=10mm,bburx=180mm,bbury=154mm,%
  width=80mm,angle=0]{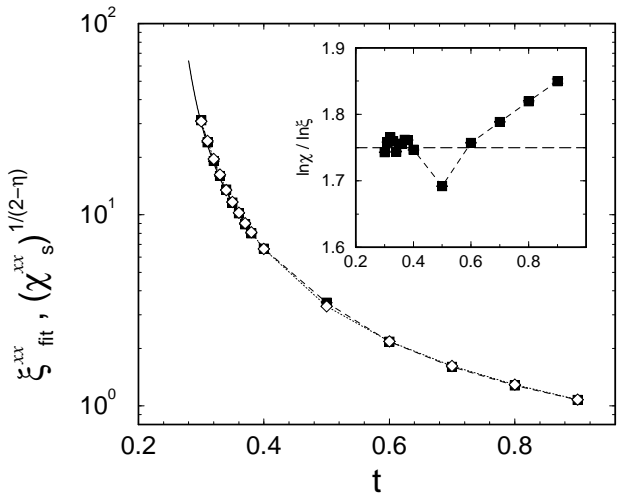}
 \caption{\label{f.EP-BKTxichi}
 Critical behaviour of $\xi^{xx}_{_{\rm fit}}$ (full squares)
 compared to that of $(\chi^{xx}_{_{\rm{s}}})^{1/(2-\eta)}$
 with $\eta = 1/4$ (open diamonds)
 for the EP model with $\Delta_\lambda=0.02$. The solid
 line is the BKT fit to the correlation length data.
 Inset: plot of $\ln\chi_{_{\rm{s}}}/\ln\xi$; the dashed line
 represents the expected BKT value $2-\eta=1.75$.}
\end{figure}

Furthermore, near criticality it is expected that the staggered
in-plane susceptibility is related to the in-plane correlation length
by the relation~\cite{Kosterlitz74}
\begin{equation}
 \chi^{xx}_{_{\rm{s}}} = K~(\xi^{xx})^{2-\eta}~,
\label{e.chixiscaling}
\end{equation}
where $K$ is a nonuniversal constant and $\eta=1/4$. By plotting
$\xi^{xx}_{_{\rm fit}}$ together with
$(\chi^{xx}_{_{\rm{s}}})^{1/(2-\eta)}$, as done in
Fig.~\ref{f.EP-BKTxichi}, one observes that this prediction also holds
for the weakly-anisotropic quantum model; remarkably, the curve
$(\chi^{xx}_{_{\rm{s}}})^{1/(2-\eta)}$ collapses onto the $\xi^{xx}$ on
a wide range of temperature so that $K\approx1$: this property is not
shared by, e.g., the classical planar rotator model or by the 2D
quantum XY model. A closer look to the validity of the scaling
relation~\eqref{e.chixiscaling} can be obtained by plotting the ratio
\begin{equation}
 \frac{\ln\chi^{xx}_{_{\rm{s}}}}{\ln\xi^{xx}}
 =2-\eta+\frac{\ln K}{\ln a_\xi+b_\xi(t-t_{_{\rm{BKT}}})^{-1/2}}~,
\end{equation}
which converges to the value $2-\eta=1.75$ when
$t\to{t_{_{\rm{BKT}}}^+}$; this is clearly shown by the data plotted in
the inset of Fig.~\ref{f.EP-BKTxichi}.

In the $\Delta_\lambda=0.001$ case, neither $\xi^{xx}_2$ nor
$\chi^{xx}_{_{\rm{s}}}$ exhibit the expected BKT critical behaviour for
the considered lattice sizes.  The correlation length obtained by
fitting the correlator $C^{xx}$ to the function~\eqref{isocorr} is also
not of much help.  This suggests the in-plane correlation length to
behave as in the isotropic model up to relatively large values
($\xi^{xx}\approx100$), and the same holds for the staggered
susceptibility. Such findings closely resemble those of neutron
scattering experiments on very weakly anisotropic layered $S=1/2$
compounds, such as Sr$_2$CuO$_2$Cl$_2$~\cite{Grevenetal95},
La$_2$CuO$_4$~\cite{Birgeneauetal99} and
Pr$_2$CuO$_4$~\cite{Sumarlinetal95}, that do not show signature of the
existing anisotropy in the correlation length and static structure
factor data.

Both the out-of-plane staggered susceptibility and correlation length
have a non-critical behaviour, with a maximum well above the
transition, at a temperature which roughly coincides with that of the
minimum of the out-of-plane uniform susceptibility, marking the onset
of an anisotropy-dominated regime. Both maxima are clearly decoupled
from the transition temperature, at variance with the maximum of the
transverse staggered susceptibility and correlation length observed in
the EA case. To this respect we mention a definite disagreement with
Ref.~\onlinecite{Ding92-EP}, where out-of-plane correlation length is
conjectured to diverge exponentially when $T\to0$. We show such
conjecture to be wrong, as $\xi^{zz}$ is clearly seen to saturate to a
finite value.

\section{Phase diagram}
\label{s.phasediag}

\begin{figure}
\includegraphics[bbllx=3mm,bblly=10mm,bburx=180mm,bbury=154mm,%
  width=80mm,angle=0]{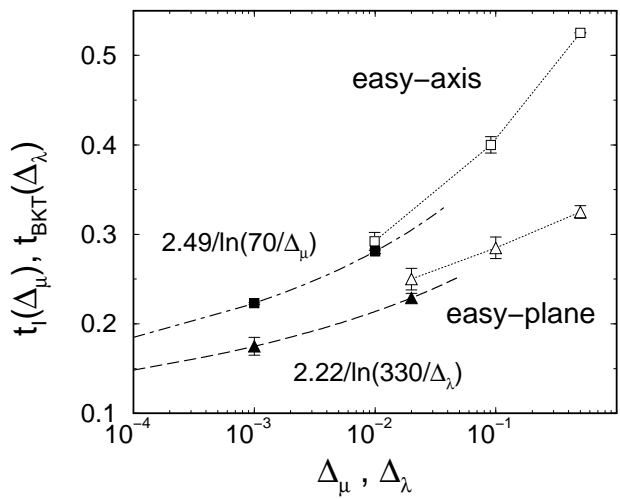}
 \caption{\label{EAEP-phdiagr+Ding}
Phase diagram of the $S=1/2$ 2$d$ XXZ model on the square lattice for
weak anisotropies. Full symbols are results of this work, open symbols
are QMC data from Refs.~\onlinecite{QMC-EA}
and~\onlinecite{Ding92-EP}.}
 \end{figure}

The detailed analysis presented above for the EA and EP models
separately, is now composed to form the phase diagram
$t_{_{\rm{I,BKT}}}$ vs $\Delta_{\mu,\lambda}$ in
Fig.~\ref{EAEP-phdiagr+Ding}, where our best estimates for the critical
temperatures relative to the four models considered are shown, together
with data from Refs.~\onlinecite{QMC-EA} and~\onlinecite{Ding92-EP}.
Critical temperatures are seen to be strongly reduced with respect to
the classical values, as given for instance in
Ref.~\onlinecite{CRTVV01}: however, the diagram clearly suggests the
critical temperatures to stay finite for any finite anisotropy, both in
the EA and in the EP case, thus leading to the conclusion that quantum
fluctuations cannot destroy the transition.

We can actually see that the above conclusion is the consequence of a
more general finding. If one numerically analyses
$t_{_{\rm{I,BKT}}}(\Delta_{\mu,\lambda})$ finds that a logarithmic
dependence is well consistent with our data, as shown in
Fig.~\ref{EAEP-phdiagr+Ding}. Such dependence, already predicted by
renormalization group techniques~\cite{Khokhlacev76,IrkhinK98-99}, is
rederived in Appendix~\ref{a.sketchy} on the basis of a fully classical
argument. It is found that
\begin{equation}
  T_{_{\rm{I}}} \approx
  \frac{4\pi\rho_{_{\rm{S}}}}{\ln(c/\Delta_\mu)}~,
\label{Isingclassical}
\end{equation}
and
\begin{equation}
  T_{_{\rm{BKT}}} \approx
  \frac{4\pi\rho_{_{\rm{S}}}}{\ln(c'/\Delta_\lambda)}~.
\label{BKTclassical}
\end{equation}
where $c$ and $c'$ are constants,
 while $\rho_{_{\rm{S}}}$ is the spin stiffness of the
classical isotropic model, entering the above expressions via the
exponential divergence of its correlation length. The dominant effect
of quantum fluctuations on such divergence is embodied in the spin
stiffness renormalization; therefore, if the ordering process we are
here observing is the same as in the classical case, we expect
$4\pi\rho_{_{\rm{S}}}=2.26\,J$, where the value
$\rho_{_{\rm{S}}}=0.18\,J$ has been taken for the renormalized  $S=1/2$
isotropic spin stiffness\cite{Beardetal98}. From the logarithmic fits
of the quantum data we indeed find $2.22$ and $2.49$ as prefactors of
the logarithm, which are remarkably near to the predicted value,
despite the simplicity of the argument that led to it.

For $\Delta_\lambda=0.02$ and $\Delta_\mu=0.01$, where a direct
comparison is possible, the critical temperatures are not fully
consistent with the values given in Refs.~\onlinecite{QMC-EA}
and~\onlinecite{Ding92-EP}.  We notice that the latter were estimated
as free parameters of fitting functions for the critical behaviour of
the susceptibility and correlation length; the precision of this
approach is hindered by the fact that the critical regime of both
quantities was not always properly entered in the simulations of
Refs.~\onlinecite{QMC-EA} and~\onlinecite {Ding92-EP}, mainly due to
technical limitations which are nowadays overcome. We therefore propose
our data as more precise estimates.

\section{Conclusions}
\label{s.conclusions}

In this paper we have presented an extensive numerical study of
thermodynamic and critical properties of weakly anisotropic
two-dimensional quantum antiferromagnets described by the 2D $S=1/2$
XXZ model with both EA and EP anisotropy. Use has been made of the
continuous-time QMC method based on the loop algorithm, implemented
here for the first time also in the EA case, and on the worm algorithm,
here reformulated as a variant of the loop algorithm.

The general outcome of the numerical simulations is that the
thermodynamics of 2D quantum antiferromagnets is extremely sensitive to
the presence of anisotropies of magnitude comparable to those of real
compounds, i.e., as small as $10^{-3}$ times the dominant isotropic
coupling.

In the models studied we see a finite temperature transition to persist
with clear signatures of Ising and BKT critical behaviour, in the EA
and EP case, respectively; in the more anisotropic case ($10^{-2}$)
full consistency with the expected universality class is found.  The
most striking evidences of the presence of the exchange anisotropy are
observed in the thermodynamic behaviour of correlation lengths and
susceptibilities. Moreover, the dependence of the critical temperature
on the anisotropy is found to be quantitatively consistent with the
prediction relative to the classical case, with properly renormalized
parameters. This tells us that quantum effects can neither destroy the
transition, nor change the ordering mechanism responsible for the
transition to occur and that our quantum models, despite having $S=1/2$
and very weak anisotropies, do actually behave as renormalized
classical ones. Given the results of Ref.~\onlinecite{CRTVV01} for
$S\ge 1$, we can say that this conclusion generally holds for quantum
Heisenberg antiferromagnets on the square lattice.

As for the thermodynamic behaviour of the specific observables
considered, we find all the non-diverging quantities to be highly
sensitive to the anisotropy, while critical quantities show deviations,
with respect to the isotropic case, which are generally harder to
detect. This is due to the fact that, in order to discriminate between
$T=0$ isotropic and finite-$T$ anisotropic divergences one must come
very close to the critical point of the anisotropic model, which is a
non-trivial issue both numerically (due to severe finite-size effects)
and experimentally (due to finite experimental resolution and
intralayer coupling).

As for the EP case, we underline that the considered values of
anisotropy compare to that of several real compounds.  On the other
hand, we have clearly shown, for instance in Figs.~\ref{EP-chi0},
\ref{EP-xi} and~\ref{EP-chi}, that in the EP case traces of 2D
anisotropic behaviour are detectable above the transition, due to the
fact that some quantities display either a minimum or a maximum in a
temperature region well apart from $t_{_{\rm{BKT}}}$, where the
in-plane correlation length has not diverged yet, and experimental
observation should hence be more feasible. We therefore think that our
results could constitute a sound basis for a possible experimental
observation of genuinely 2D EP behaviour in real magnets.

\begin{acknowledgments}
The numerical calculations which led to the present work have been
performed on the parallel cluster at CILEA (Milano, Italy), on the
beowulf cluster CINECA (Bologna, Italy) through INFM grant
no.\,1114174655, on the beowulf cluster at INOA (Firenze, Italy) and on
the PC's of the Instit\"ut f\"ur theoretische Physik at the University
of Leipzig (Germany). We thank all these institutions for their
generous support. We thank T.~Sauer and W.~Janke for participation to
the early stage of this project. T.R. acknowledges fruitful
conversations with B.B.~Beard, N.~Kawashima, P.~Carretta and
M.~Laurati. T.R. further acknowledges the Instit\"ut f\"ur theoretische
Physik at Leipzig University for hospitality during the early steps of
this project, and CRUI for financial support via the VIGONI programme.
This work has been partially supported by the COFIN2000-MURST fund.

\end{acknowledgments}

\appendix

\section{Ising limit of the worm algorithm}
\label{a.worm}

We show that the Ising limit of the worm algorithm is nothing but the
standard single-flip Metropolis algorithm, considering the simple case
of an Ising antiferromagnetic spin dimer. Independent transitions via
single spin flip are only two: ({\it{i}})
$|{\uparrow\uparrow}\rangle\to|{\downarrow\uparrow}\rangle$ and
({\it{ii}}) its reversal. Within the worm algorithm, the transition
({\it{i}}) occurs with probability 1, since the spin configuration
$|{\uparrow\uparrow}\rangle$ admits no bounce; this equals the
Metropolis result, since a net energy gain $J^Z/2$ is obtained in the
process ({\it{i}}). Starting from the configuration
$|{\downarrow\uparrow}\rangle$ a bounce process is instead possible
with a probability per unit time $\pi(3,b)=J^Z/2$; if the worm
experiences even a single bounce, it will u-turn finding on its way
back only plaquettes of type 1(2), on which it cannot bounce, until it
reaches its tail leaving the configuration unchanged. Therefore the
transition occurs with probability $1-p_{\rm 1b}$, where $p_{\rm 1b}$
is the probability of experiencing at least one bounce. Since the
number of bounces follows a Poisson distribution with parameter
$\beta\pi(3,b)$, we have that
\begin{equation}
 1-p_{\rm{1b}}= 1-\sum_{k\geq1} \frac{[\beta\pi(3,b)]^k}{k!}
 ~e^{-\beta \pi(3,b)} = e^{-\beta J^Z/2}
\end{equation}
which is exactly the Metropolis transition probability for a process
involving an energy increase $J^Z/2$.

\section{Estimator of the helicity modulus}
\label{a.helicity}

In this section we derive the QMC estimator~\eqref{winding} for the
helicity modulus starting from its thermodynamic
definition~\eqref{thermohelix}.  The derivation is a finite-temperature
generalization of the one given by Sandvik in
Ref.~\onlinecite{Sandvik97}, in the context of Stochastic Series
Expansion, to estimate the spin stiffness, i.e., at zero temperature.

We start from the "twisted" XXZ hamiltonian, with the twist applied
along the 1-direction of the lattice, as:
\begin{eqnarray}
 \hat{\cal H}(\phi) &=& \sum_{\bm{i}}
 \big[J^{XY}\cos\phi(\hat{S}_{\bm{i}}^x\hat{S}_{{\bm{i}}+{\bm{d_1}}}^x
 + \hat{S}_{\bm{i}}^y\hat{S}_{{\bm{i}}+{\bm{d_1}}}^y)
\\ \nonumber
 &+& J^{XY}\sin\phi(\hat{S}_{\bm{i}}^x\hat{S}_{{\bm{i}}+{\bm{d_1}}}^y-
 \hat{S}_{\bm{i}}^y\hat{S}_{{\bm{i}}+{\bm{d_1}}}^x)
 +  J^Z \hat{S}_{\bm{i}}^z\hat{S}_{{\bm{i}}+{\bm{d_1}}}^z \big]
\\ \nonumber
 &+& {\hat{\cal H}}_2
\label{twist}
\end{eqnarray}
where ${\bm{d_1}}=(1,0)$ and ${\hat{\cal H}}_2$ is the term containing
only bonds along the 2-direction, which remains unchanged.

We expand the twisted Hamiltonian to second order in $\phi$ as:
\begin{equation}
 \hat{\cal H}(\phi) = \hat{\cal H}(\phi=0) -
 \phi \hat{\cal J}_1 - \frac{\phi^2}{2}\hat{\cal H}^{(XY)}_1 +
 {\cal O}(\phi^3)
\end{equation}
where
\begin{equation}
 \hat{\cal J}_1 = \frac{iJ^{XY}}{2}\sum_{\bm{i}}
 \big(\hat{S}_{\bm{i}}^+\hat{S}_{{\bm{i}}+{\bm{d_1}}}^- -
 \hat{S}_{\bm{i}}^-\hat{S}_{{\bm{i}}+{\bm{d_1}}}^+\big)
\end{equation}
is the $1$-component of the spin current operator, and
\begin{equation}
 \hat{\cal H}^{(XY)}_1= \frac{J^{XY}}{2}
 \sum_{\bm{i}}\big(\hat{S}_{\bm{i}}^+\hat{S}_{{\bm{i}}+{\bm{d_1}}}^- +
 \hat{S}_{\bm{i}}^-\hat{S}_{{\bm{i}}+{\bm{d_1}}}^+\big)~.
\end{equation}

Carefully deriving the free energy with respect to the twist, i.e.,
taking care of the non-commutativity between the $\hat{\cal
H}(\phi=0)$, $\hat{\cal J}_1$ and $\hat{\cal H}^{(XY)}_1$ operator, one
obtains for the helicity modulus, averaged over the $1$- and
$2-$direction of the applied twist, the expression:
\begin{equation}
 \Upsilon = - \frac{1}{2J^{XY}L^2}
 \bigg(\big\langle\hat{\cal H}^{(XY)}\big\rangle
 +  \int_{0}^{\beta} d\tau \big\langle \hat{\bm J}(0)
 \cdot \hat{\bm J}(\tau)\big\rangle \bigg)
\label{helicity}
\end{equation}
where $\hat{\cal H}^{(XY)}= \hat{\cal H}^{(XY)}_1 +
\hat{\cal H}^{(XY)}_2$ and
$\hat{\bm J}=(\hat{\cal J}_1,\hat{\cal J}_2)$. Such expression stands
as the direct quantum analogue of the classical expression as given in,
{\it e.g.}, Ref.~\onlinecite{SchultkaM94} for the plane rotator model;
in the limit of zero temperature it reproduces the expression given by
Ref.~\onlinecite{Sandvik97}.

The QMC estimator~\eqref{winding} can be obtained directly starting
from~\eqref{helicity} in the case $S=1/2$. In the continuous-time
limit, the estimator for the bond exchange operators
$\hat{T}^{\pm}_{{\bm{i}}{\bm{d_1}}} = (J^{XY}/2) ~
\hat{S}_{\bm{i}}^{\pm}\hat{S}_{{\bm{i}}+{\bm{d_1}}}^{\mp}$ takes the
form:
\begin{equation}
 \big\langle\hat{T}^{\pm}_{{\bm{i}}{\bm{d_1}}}\big\rangle =
 \frac{1}{\beta} \int_0^{\beta} d\tau
 \big\langle\hat{T}^{\pm}_{{\bm{i}}{\bm{d_1}}}(\tau)\big\rangle =
 -\frac1\beta\big\langle N^{\pm}_{{\bm{i}}{\bm{d_1}}}
 \big\rangle_{_{\rm{MC}}}
\end{equation}
where $N^{+}_{{\bm{i}}{\bm{d_1}}}$ is the number of (+)-kinks (of the
type \mbox{$|{\downarrow_{\bm{i}}\uparrow_{{\bm{i}}+{\bm{d_1}}}}\rangle
\to|{\uparrow_{\bm{i}}\downarrow_{{\bm{i}}+{\bm{d_1}}}}\rangle$})
on the ${{\bm{i}},{\bm{i}+d_1}}$ bond, and analogously for
$N^{-}_{{\bm{i}}{\bm{d_1}}}$. Therefore the estimator for the XY-energy
takes the form:
\begin{equation}
 \big\langle \hat{\cal H}^{(XY)}\big\rangle =
 -\frac{1}{\beta}\big\langle N^{+}+N^{-}\big\rangle_{_{\rm{MC}}}
\label{enestimator}
\end{equation}
where $N^{\pm}$ is the total number of ($\pm$)-kinks present in each
configuration. The current-current correlator present
in~\eqref{helicity} can be decomposed into bond-pair contributions as
follows:
\begin{equation}
 \hat{\cal J}_1(0) \hat{\cal J}_1(\tau)
 = \!-\! \sum_{{\bm{i}},{\bm{j}}}
 \big[\hat{T}^{+}_{{\bm{i}}{\bm{d_1}}}(0)
 -\hat{T}^{-}_{{\bm{i}}{\bm{d_1}}}(0)\big]
 \big[\hat{T}^{+}_{{\bm{j}}{\bm d_1}}(\tau)
 -\hat{T}^{-}_{{\bm{j}}{\bm{d_1}}}(\tau)\big]
\end{equation}
Taking into account the $S{=}1/2$ constraint
$\hat{S}^{\pm}\hat{S}^{\pm}|\sigma\rangle=0$, one obtains that
\begin{eqnarray}
 &&\int_{0}^{\beta} d\tau
 \Big\langle\big[\hat{T}^{+}_{{\bm{i}}{\bm{d_1}}}(0)-
 \hat{T}^{-}_{{\bm{i}}{\bm{d_1}}}(0)\big]
 \big[\hat{T}^{+}_{{\bm{j}}{\bm{d_1}}}(\tau)
 -\hat{T}^{-}_{{\bm{j}}{\bm{d_1}}}(\tau)\big]\Big\rangle=
\nonumber \\
 & &~~ =  \frac{1}{\beta} ~ \bigg[~\Big\langle
 \big(N^{+}_{{\bm{i}}{\bm{d_1}}}- N^{-}_{{\bm{i}}{\bm{d_1}}}\big)
 \big(N^{+}_{{\bm{j}}{\bm{d_1}}}-N^{-}_{{\bm{j}}{\bm{d_1}}}\big)
 \Big\rangle_{_{\rm MC}}
\nonumber \\
 && \hspace{28mm}-~\delta_{{\bm{i}}{\bm{j}}}~
 \Big\langle\big(N^{+}_{{\bm{i}}{\bm{d_1}}}+
 N^{-}_{{\bm{i}}{\bm{d_1}}}\big) \Big\rangle_{_{\rm MC}} ~\bigg]
\label{correstimator}
\end{eqnarray}
Putting together~\eqref{enestimator} and~\eqref{correstimator}
with~\eqref{helicity} one obtains:
\begin{equation}
\Upsilon = \frac{t}{2L^2}
\left\langle \left(N^{+}_1-N^{-}_1\right)^2 +
\left(N^{+}_2-N^{-}_2\right)^2\right\rangle_{_{\rm MC}}
\label{helicity2}
\end{equation}
where $(N^{\pm}_{\rm 1(2)}$ is the total number of $\pm$-kinks on
1(2)-bonds. Since a (+)-kink and a ($-$)-kink affect a spin path
crossing the kink by shifting it of a lattice spacing in opposite
directions, we have that the spin-path winding number can be expressed
as
\begin{equation}
 W_{\rm 1(2)} = \Big( N^{+}_{\rm 1(2)}-N^{-}_{\rm 1(2)}\Big) \big/ L
\end{equation}
and this reduces the estimator~\eqref{helicity2} to the
expression~\eqref{winding}.

\section{Classical description of the ordering mechanism}
\label{a.sketchy}

We give here a sketchy description of the ordering mechanism in
slightly anisotropic 2D magnets, referring to the classical limit where
the antiferro- and ferromagnetic cases are thermodynamically
equivalent.

{\em EA case}.~ We rewrite the Hamiltonian~\eqref{e.xxzmodel} in the
classical limit as
\begin{eqnarray}
  {\cal H} &=& -\frac{J_{\rm{cl}}}2~ \sum_{{\bm{i}},{\bm{d}}}
  {\bm s}_{\bm{i}} \cdot {\bm s}_{{\bm{i}}+{\bm{d}}} ~+~ {\cal H}'
  \nonumber \\
  {\cal H}' &=& \frac{J_{\rm{cl}}\Delta_\mu}2~\sum_{{\bm{i}},{\bm{d}}}
  ( s_{\bm{i}}^x s_{{\bm{i}}+{\bm{d}}}^x+ s_{\bm{i}}^y s_{{\bm{i}}+{\bm{d}}}^y)
\end{eqnarray}
where ${\bm s}=(\cos\theta\sin\phi,\sin\theta\sin\phi,\cos\theta)$ is a
unitary classical vector and $J_{\rm{cl}}$ is the classical exchange
constant. In the above form the Hamiltonian is written as the isotropic
Heisenberg Hamiltonian plus, as long as $\Delta_\mu\ll{1}$, a small
Ising perturbation ${\cal{H}}'$. The isotropic Heisenberg model has an
exponentially divergent correlation length as
$T\to0$~\cite{BrezinZ-J76-CaraccioloP94},
$\xi\approx{aT}e^{2\pi\rho_{_{\rm{S}}}/T}$, where $\rho_{_{\rm{S}}}$ is
the spin stiffness of the classical model. At very high temperatures,
i.e., $T\gg{J_{\rm{cl}}}$, the spins are fully uncorrelated, so that
the anisotropy has little effect. When $T\approx{J_{\rm{cl}}}$
correlations set on and clusters of almost aligned spins form on the
length scale $\xi$. Very roughly, one can imagine the $\xi^2$ spins of
each cluster $C$ to be fully aligned, so that the anisotropy term can
be written as
\begin{eqnarray}
  {\cal H'}  & = & \frac{J_{\rm{cl}}\Delta_\mu}2 \sum_{{\bm{i}},{\bm{d}}}
  \cos(\phi_{\bm{i}}-\phi_{{\bm{i}}+{\bm{d}}})
  \sin\theta_{\bm{i}}~\sin\theta_{{\bm{i}}+{\bm{d}}}
\nonumber \\
  &\approx& J_{\rm{cl}}\Delta_\mu\,\xi^2\, \sum_C\sin^2\theta_C~,
\end{eqnarray}
where $\theta_C$ is the polar angle of the spin orientation on each
cluster and border terms of order $\xi$ are neglected. Hence, the
anisotropy term creates an effective potential for the orientation of
each correlated cluster that has two minima in $\theta_C=0$ and $\pi$
(up and down) separated by an energy barrier
$\Delta{E}=J_{\rm{cl}}\Delta_\mu\,\xi^2$. When $\xi$ increases upon
lowering $T$, the barrier becomes comparable to the thermal energy, so
that the orientation of each cluster is confined to one side of the
potential barrier: the system becomes Ising-like and a finite
magnetization appears. Using the isotropic behaviour of $\xi$, this
happens when
\begin{equation}
  T \approx \Delta E = J_{\rm{cl}}\Delta_\mu
  \big(aT\,e^{2\pi\rho_{_{\rm{S}}}/T}\big)^2 ~;
\end{equation}
solving this equation gives the critical temperature as in
Eq.~\eqref{Isingclassical}.

The above simplified picture accounts for the logarithmic dependence of
$T_{\rm{c}}$ upon the (small) anisotropy $\Delta_\mu$ that was earlier
obtained via more sophisticated approaches~\cite{IrkhinK97}; the
logarithm appears to follow from the exponential correlation length in
the isotropic model.

{\em EP case}.~ The above argument can be essentially rephrased, this
time taking as perturbation of the isotropic Hamiltonian the term
\begin{equation}
{\cal H'} = \frac{J_{\rm{cl}}\Delta_\lambda}2~\sum_{{\bm{i}},{\bm{d}}}
  s_{\bm{i}}^z s_{{\bm{i}}+{\bm{d}}}^z ~,
\end{equation}
which, in presence of clusters on the scale $\xi$, becomes
\begin{equation}
  {\cal H'} \approx J_{\rm{cl}}\Delta_\lambda\,\xi^2\,
            \sum_C \cos^2\theta_C~.
\end{equation}
The anisotropy potential has the minimum at $\theta_C=\pi/2$, i.e., for
a cluster orientation on the $xy$-plane, and the well depth is
$\Delta{E}=J_{\rm{cl}}\Delta_\lambda\,\xi^2$. As in the EA case, the
anisotropy becomes relevant once $T$ is comparable to $\Delta{E}$: the
out-of-plane fluctuations are suppressed making the system effectively
planar, so that vortex excitations appear and the BKT transition can
take place. This roughly happens when Eq.~\eqref{BKTclassical} holds.


\begin{thebibliography}{200}
\bibitem{nota}
The weigth of plaquette configurations of type 1 (2) differ from those
given in Eq.(36) of Ref.~\onlinecite{Evertz} as we did not multiply by
the arbitrary factor $\exp(-\frac{J^Z}{4}\Delta\tau)$; this also
affects the break-up weights given in Eqs.~\eqref{EP-breakup}
and~\eqref{EA-breakup}.
\end{thebibliography}
\end{document}